\documentclass{emulateapj}
\usepackage{natbib}
\def \ergpers {erg~s$^{-1}$}
\def \gevperspercm {GeV~s$^{-1}$~cm$^{-2}$}

\slugcomment{Draft version \today}

\shorttitle{Propagation of ultrahigh energy nuclei in clusters of galaxies}
\shortauthors{Kotera et al.}

\begin{document}


\title{Propagation of ultrahigh energy nuclei in clusters of galaxies: \\resulting composition and secondary emissions}


\author{K. Kotera\altaffilmark{1}, D. Allard\altaffilmark{2}, K. Murase \altaffilmark{3},
          J. Aoi\altaffilmark{3},
          Y. Dubois\altaffilmark{4},
          T. Pierog\altaffilmark{5}}
          \author{
          S. Nagataki\altaffilmark{3}
}
\email{kotera@iap.fr}


\altaffiltext{1}{Institut d'Astrophysique de Paris, UMR7095 - CNRS, Universit\'e Pierre \& Marie Curie, 98 bis boulevard Arago, F-75014 Paris, France.}
\altaffiltext{2}{Laboratoire Astroparticules et Cosmologie,  (APC), Universit\'e Paris 7/CNRS, 10 rue A. Domon et L. Duquet, 75205 Paris Cedex 13, France.}
\altaffiltext{3}{Yukawa Institute for Theoretical Physics, Kyoto University, Kyoto 606-8502, Japan.}
\altaffiltext{4}{University of Oxford, Astrophysics, Denys Wilkinson Building, Keble Road, Oxford OX1 3RH.}
\altaffiltext{5}{Forschungszentrum Karlsruhe, Institut fuer Kernphysik, Karlsruhe, Germany.}


\begin{abstract}

We study the survival of ultrahigh energy nuclei injected in clusters of galaxies, as well as their secondary neutrino and photon emissions, using a complete numerical propagation method and a realistic modeling of the magnetic, baryonic and photonic backgrounds. It is found that the survival of heavy nuclei highly depends on the injection position and on the profile of the magnetic field. Taking into account the limited lifetime of the central source could also lead in some cases to the detection of a cosmic ray afterglow, temporally decorrelated from neutrino and gamma ray emissions. 

We calculate that the diffusive neutrino flux around 1~PeV coming from clusters of galaxies may have a chance to be detected by current instruments. The observation of single sources in neutrinos and in gamma rays produced by ultrahigh energy cosmic rays will be more difficult. Signals coming from lower energy cosmic rays ($E\lesssim 1$~PeV), if they exist, might however be detected by Fermi, for reasonable sets of parameters. 

\end{abstract}







\section{Introduction}
 
The origin of ultrahigh energy cosmic rays still remains an unsolved puzzle. Most puzzling is the fact that at the highest energies, the cosmic ray arrival directions as observed by the current and previous experiments do not point back to any of their powerful candidate sources. One may underline that the Pierre Auger Observatory has reported a significant correlation between the arrival direction of cosmic rays of energy $E> 5.7\times 10^{19}$~eV and a catalogue of Active Galactic Nuclei (AGN) located at less than $75$~Mpc \cite[]{Auger1}. This correlation however is not confirmed and should be interpreted as an evidence of an anisotropic distribution of sources in the sky (\citealp{Auger09}, and see for example \citealp{KL08b}, \citealp{KW08}, \citealp{Ghisellini08}, \citealp{Cuo+08}, \citealp{Tak+08}, \citealp{ZFG09} for interpretations of these results). 

Magnetic fields certainly play a key role in this puzzle. Yet our lack of knowledge concerning the distribution and strength of these fields, at both Galactic and extragalactic scales, introduces a large degree of uncertainty in the subject of charged particle propagation. 

In view of this picture, one promising way to identify the sources of ultrahigh energy cosmic rays is to resort to multimessenger Astronomy. If charged particles are subject to deflections from magnetic fields, the direction of secondary photons and neutrinos\footnote{In the following, we will refer to photons and neutrinos produced by interactions of cosmic rays with the ambient medium as \textit{secondary} photons and neutrinos.} 
they produce by interaction with the ambient medium should be much less affected during their propagation. 

Observing such secondary particles by using current detectors is not a straightforward  task. Even though the existence of cosmogenic neutrinos is guaranteed, their detectability by present and future experiments is not. Indeed, their flux crucially depends on astrophysical parameters such as the cosmological evolution of the cosmic ray luminosity, the maximum energy at the sources or the chemical composition \cite[]{Allard06, Takami09}. As for secondary high energy photons, many studies have addressed the question of their detection \cite[]{Ferrigno04, GA05,GA07, ASM06}, but the situation remains unclear. These works indeed conclude that the observation of these photons strongly depends on assumptions on the configuration and strength of the intergalactic magnetic fields, as well as on the source luminosity in cosmic rays.\\

Clusters of galaxies can be considered in various means as an ideal secondary particle production region. First they are dense regions of the Universe that harbour many candidate sources for ultrahigh energy cosmic rays, such as AGN, compact stellar remnants and cosmological shocks. We will discuss the importance of these possible accelerators in our framework in section~\ref{subsection:sources}. Second, clusters of galaxies have enhanced photonic and baryonic backgrounds as compared to the extra-cluster medium. Indeed, X-ray observations of the bremsstrahlung emission in clusters of galaxies have revealed that a large fraction of the cosmological baryonic density was contained in the intracluster gas \cite[]{WF95}. 
Besides, the high concentration of galaxies produces a local overdensity of infrared photons \cite[]{LPD05}. 

The most outstanding feature of clusters of galaxies for high energy cosmic ray propagation is certainly their strong magnetisation. 
Clusters of galaxies are the only extragalactic regions where magnetic fields could actually be observed, by measuring the Faraday rotation of linearly polarised emission. They can reach values ranging from a few microgauss on scales of order $\sim 10$~kpc for normal clusters \cite[]{Kim91, CKB01, Clarke04}, up to $10 - 40~\mu$G on scales of $3 - 5$~kpc for cool core clusters \cite[]{TP93,EV06}. 
Such strong magnetic fields can easily confine cosmic rays of energy $E\lesssim 5\times 10^{17}~\mbox{eV}\times Z$,  where $Z$ is the electric charge, for several hundreds of millions of years, leading galaxy clusters to act like storage rooms for these particles. While propagating inside the cluster, cosmic rays can thus experience many interactions with the enhanced photonic and baryonic backgrounds, and produce secondary neutrinos and photons that might be detectable with current and upcoming instruments. 

Many analytical and semi-analytical works were conducted on this subject to calculate the secondary gamma ray and neutrino fluxes, and to investigate the contribution of high energy protons produced in clusters of galaxies to the total spectrum \cite[]{DS95, DS96, BBP97, CB98, demarco06, ASM06, MIN08, Wolfe08}. Estimates of these fluxes can be calculated by assuming a proton injection spectrum, a Kolmogorov diffusion regime in the magnetic field, and a baryonic background profile. Numerical propagations of ultrahigh energy protons in more realistic three dimensional cluster magnetic fields were also studied by \cite{RGD04}. 

It can be summarised from these studies that it would be difficult to detect $\sim$PeV energy neutrinos emitted from single sources with the upcoming generation of telescopes such as IceCube and KM3Net. The cumulative emission however may have some chance to be observed, depending on the source luminosity in cosmic rays and on the background densities. GeV/TeV gamma ray signals may also be observed, again depending on assumptions on physical parameters: in particular, \cite{CB98} find that the cluster population can account for a
fraction $\sim 0.5 - 2\%$ of the diffuse gamma ray background  and derive a list of clusters that could be observable in gamma ray with the Fermi Space Telescope. \cite{ASM06} and \cite{MIN08} obtain less optimistic results, but the former authors find that a source located  at 100~Mpc with maximum proton energy  $E_{\rm max}\sim 10^{21}$~eV and injection spectral index ranging from 2.3 to 2.7 can be detected by HESS2.

One should note that these previous studies generally assume a spherical symmetry for the modelling of the galaxy cluster baryonic density and magnetic field (apart from \citealp{RGD04} who propagate particles in a simulated cluster magnetic field). The infrared photon background was taken into account only in the work by \cite{demarco06}, but in a highly optimistic way. Besides, all these studies investigate the propagation of {\it protons} inside clusters of galaxies. \\

The chemical composition of high energy cosmic rays is still an open question though. The KASCADE data and measurements prior to the Pierre Auger Observatory indicate a dominance of heavy nuclei around the knee region followed by a transition towards a lighter composition around $E\sim 10^{18.5}$~eV \cite[]{KASCADE, Fly, AGASA, Hires}. The results of the Pierre Auger Observatory tend to suggest a mixed composition at all energies, that might   even get heavier at the highest end \cite[]{U07}. 

The energy losses of nuclei with mass number $A>1$ during the propagation definitely differs from those of protons. They were first studied in detail by \cite{PSB76} and have been reexamined by \cite{SS99}, \cite{ER98} and most recently by \cite{Khan05}. The implementation of these results analytically or in detailed numerical simulations for the propagation of ultrahigh energy cosmic rays \cite[]{Bertone02,Allard05,Allard06,Allard07b} have demonstrated that iron nuclei can travel hundreds of megaparsecs before losing their energy. This interesting property combined moreover with the fact that such nuclei can be accelerated to an energy typically $Z$ times larger than protons, has motivated several authors to consider heavy nuclei as primary cosmic rays \cite[]{Anchordoqui99,Allard06,Allard07a,Allard07b,Inoue07,Anchordoqui08,Allard08}. \cite{Allard08} conclude that whatever the composition injected at the source, the propagated spectrum detected at the Earth should be either dominated by protons, or by heavy nuclei of iron type. The latter case occurs if the source is strongly enriched in iron or if the proton maximal injection energy is smaller than the GZK cut-off energy, so that only heavy nuclei are present at the greatest energies.

It is thus of prime importance to have an idea of the composition that escapes from the sources of ultrahigh energy cosmic rays, and this question should also be addressed in the context of propagation inside clusters of galaxies, considering that they might host powerful accelerators. There is no reliable prediction of the expected composition at the source, mainly because very little is known on the physical parameters that govern the acceleration and survival of nuclei in those powerful objects. As a case study nonetheless, one might reasonably consider a composition similar to the Galactic cosmic ray one, that contain around 30\% of CNO and heavier nuclei. Moreover, AGN that we will consider as sources of cosmic rays in our context, are known to have supersolar metallicities \cite[]{Groves06,Mathur09}, which further justifies our choice of injecting a mixed composition.
Because the diffusion time in a magnetic field increases with the charge $Ze$ of the particle, heavy nuclei should remain confined longer times in the structure, leading to an enhanced number of interactions and thus possibly to complete depletion of the original particle. Distinct signatures of these propagation effects could be observed in the produced spectrum as well as in the neutrino and gamma ray fluxes. \\

In the present paper, we explore the consequences of the injection of a mixed chemical composition in galaxy clusters, by calculating the propagated primary and secondary particle fluxes. We elaborated for this purpose a propagation code that treats all baryonic and photonic interaction processes for primary and secondary nuclei, as well as their accurate trajectories in the magnetic field. gamma ray signals formed by electromagnetic cascades of ultrahigh energy pairs and photons in the intergalactic space are calculated in a  second step as a post-analysis. We model clusters of galaxies using three dimensional outputs of magneto hydrodynamical (MHD) simulations run by \cite{DT08}, making a distinction between cool core and non cool core clusters. 

This paper is organized as follows: in section~\ref{section:modelling}, we will describe our physical modelling of galaxy clusters for high energy cosmic ray propagation. The details of the numerical techniques of our propagation code are given in section~\ref{section:code}. We will present and discuss our results are presented in section~\ref{section:results}.

\section{Modeling clusters of galaxies \\for ultrahigh energy cosmic ray propagation}\label{section:modelling}

We discuss in this section our modelling of clusters of galaxies for ultrahigh energy cosmic ray propagation. This comprises a three dimensional modelling of the magnetic field and infrared photon background, a consistent baryonic background profile and an adequate choice of sources for injection. Features for both cool core and non cool core clusters will be discussed. 

\begin{figure}[htb]
\begin{center}
\includegraphics[width=0.4\textwidth]{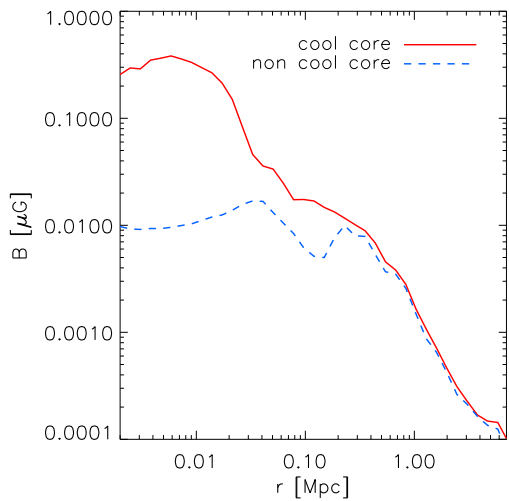} 
\caption{Magnetic field intensity profiles from the simulations of \cite{DT08}. The red solid line represents the mean magnetic field intensity in radial bins, for the cool core cluster of galaxies. The blue dashed line is for the non cool core cluster. The magnetic field intensity is rescaled for our simulations to obtain a field of 1, 3 or 10~$\mu$G at the center for cool core clusters, and 1~$\mu$G for non cool core clusters, see text for details.}  \label{fig:B}
\end{center}
\end{figure}

\begin{figure*}[htb]
\includegraphics[width=0.33\textwidth,height=0.327\textwidth]{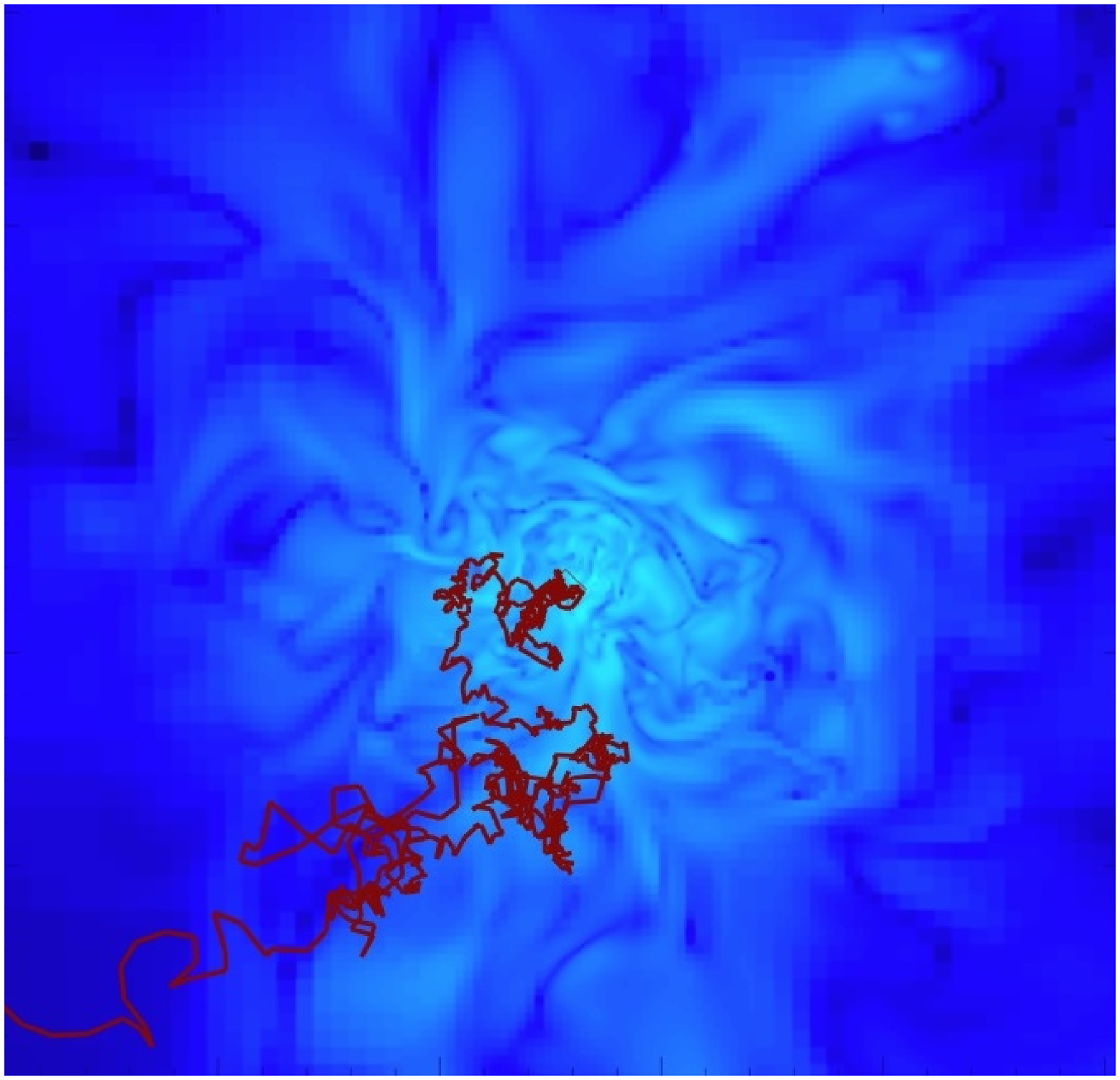} 
\includegraphics[width=0.33\textwidth,height=0.327\textwidth]{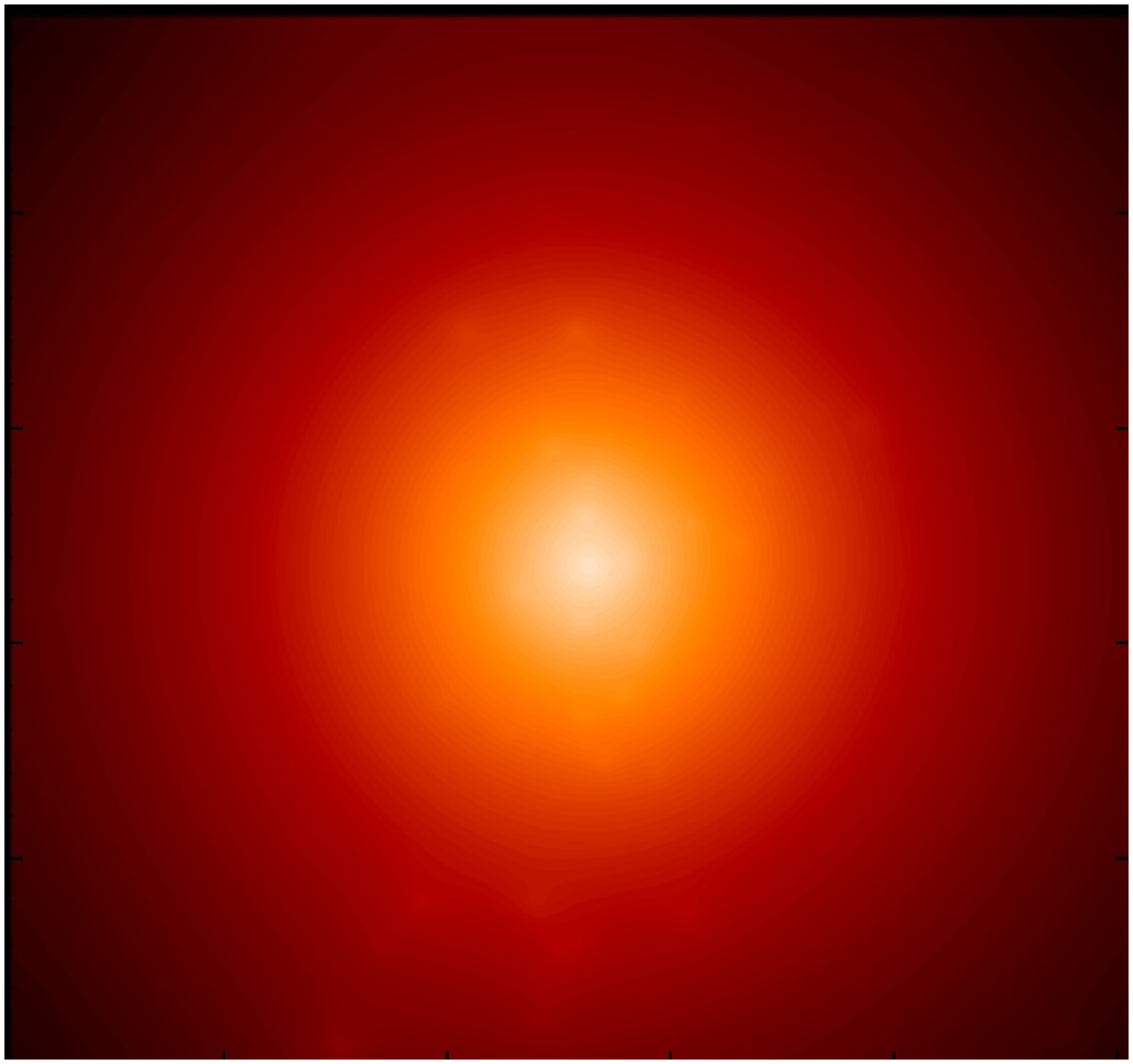}
\includegraphics[width=0.33\textwidth,height=0.328\textwidth]{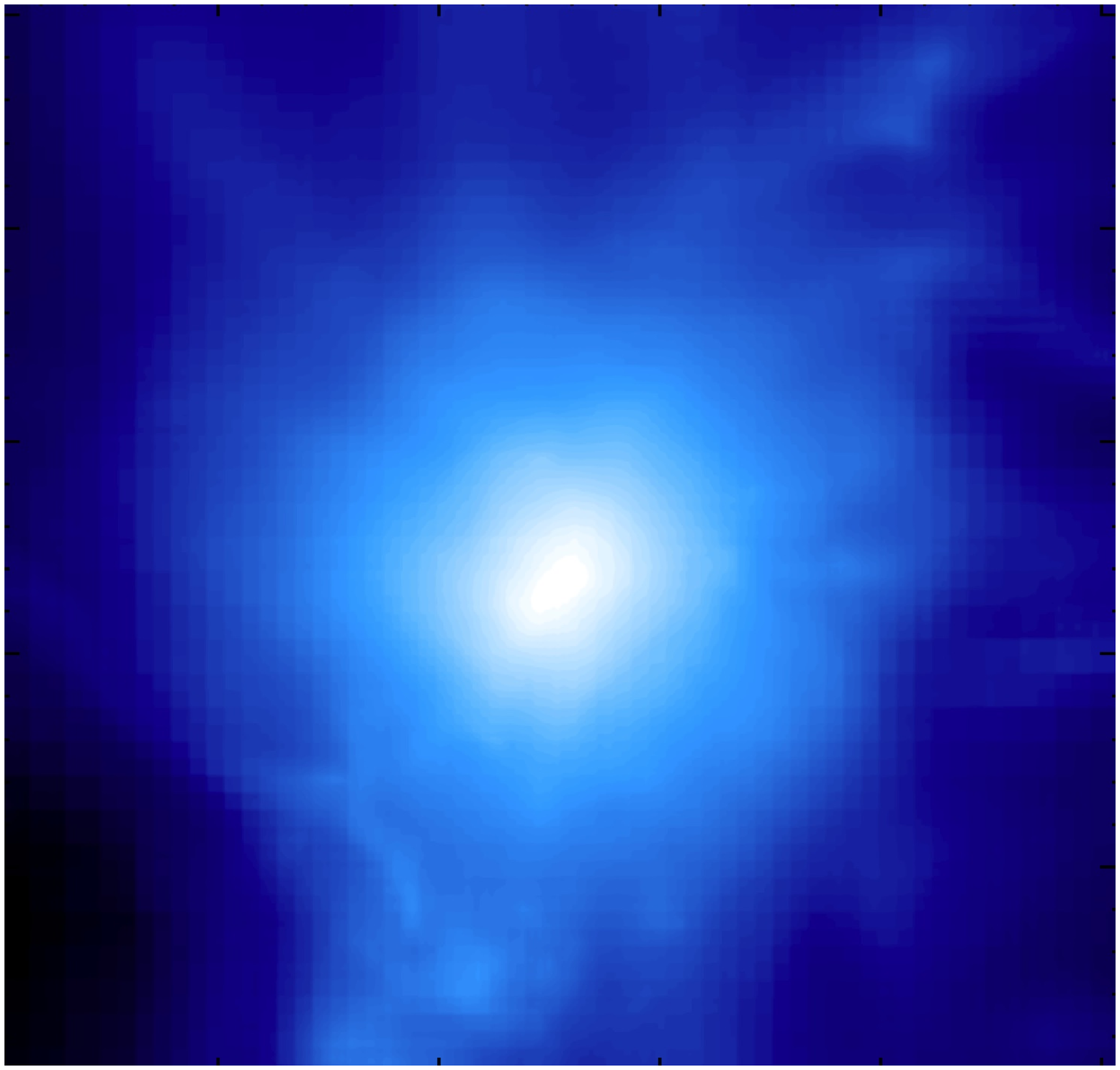}  
\caption{From left to right: magnetic field intensity, infrared photon density and baryon density projected along the observed plane, from the simulation outputs of \cite{DT08}, in the case of a cool core cluster. The colour contours are in logarithmic scale and the box size is of $\sim7$~Mpc. The red line on the magnetic field slice represents the trajectory of one proton of energy $E=10^{17}$~eV propagated in our simulations. The particle reaches the edge of the box in a time $\sim 200$~Myr. }  \label{fig:3slices}
\end{figure*}

\subsection{Magnetic field}\label{subsection:B}

Clusters of galaxies can be roughly split into two categories: cool core clusters and non cool core clusters. Observations indeed indicate that half of the cluster population exhibits a peaked X-ray emission at the core and a central temperature of 30 to 40\% of the virial temperature \cite[]{Chen07,Burns08}. This bimodality has also some significant consequences when it comes to modelling magnetic fields: cool core clusters are reported to have a much stronger magnetic field in the core, as well as a higher turbulence rate, as compared to non cool core clusters (see \citealp{CT02} for a recent review). Such differences can play a non negligible role in cosmic ray confinement, and necessitates some investigations. 

We model our cluster magnetic field using the three dimensional outputs of the MHD simulations run by \cite{DT08}. The simulations were run including Dark Matter, gas, ultraviolet heating, hydrogen and helium cooling, star formation and magnetic fields with the Adaptative Mesh Refinement code RAMSES \citep{T02}. The galaxy cluster has been evolved up to $z=0$ within a `standard' $\Lambda$CDM cosmology with parameters $\Omega_m=0.3$, $\Omega_{\Lambda}=0.7$, $\Omega_b=0.04$, $H_0=70\,  \rm km\,s^{-1}\,Mpc^{-1}$ and $\sigma_8=0.9$. The cool core cluster has the following properties at $z=0$:  $R_{200}$ (the radius inside which the mean interior overdensity is equal to 200) equals $1.1 \, h^{-1}\rm Mpc$, the virial mass $M_{200}\equiv200\times (4\pi/3)\rho_c R_{200}^3=3.5\times10^{14}\,  h^{-1}\rm M_{\odot}$ (where $\rho_c$ is the critical density), and the X-ray temperature $T_{X}=5.1 \, \rm keV$. 

Though this simulation enables one to have a very accurate resolution of $1.2 \, h^{-1}\rm kpc$ at the center of the cluster, the outputs were smoothed for practical reasons on a cubic grid of $256^3$ regular cells and of size $5\, h^{-1}\rm Mpc$, leading to a constant $19.5 \, h^{-1}\rm kpc$ resolution.
The authors performed two simulations: one including atomic cooling and star formation, which led to the formation of a cool core cluster, and the other without these processes, leading to a non cool core cluster. As a result of the increase of density due to the cooling flow, the magnetic field is stronger in the cooling run than in the adiabatic run by more than one order of magnitude, reaching a typical value of $0.5 \, \mu$G in the most favourable case (see the magnetic field radial profiles in Fig.~\ref{fig:B}). In both cases, compression and shear motions of the the hot plasma medium amplify and give rise to a non-trivial topology of the field lines at large scales (see Fig.~\ref{fig:3slices}). Thus, it is of particular interest to propagate particles in these realistic magnetic fields that can strongly vary on kpc scales (see the typical coherence lengths on Tab.~\ref{table:lcoh}). 

Many studies indicate that AGN jets can grow into large cavities through the hot plasma of the ICM \citep{AFEJF84}. \cite{BT95} show that those high velocity jets could disrupt the formation of cooling flows in the core. This is a popular explanation to the non detection by Chandra and XMM of line emissions of cold gas that would have been expected in a standard cooling flow picture at cluster cores. Such re-heating processes are not included in the work of \cite{DT08}, but drawing a complete picture of a self-consistently growing supermassive black-hole powering a high-velocity jet including magnetic fields still remains a challenge in numerical treatment. 

\cite{DT08} scale their magnetic field seeds according to previous works by \cite{DGST05}: the initial comoving magnetic field $B_0=B(z) / (1+z)^{2}$ was set up to an arbitrary low value and then renormalized to $10^{-11} $G. It can be noted however that these authors consider non cool core clusters. It is thus justified to perform a new normalization of the field in adequacy to the type of the cluster that we study -- up to the point where magnetic fields become dominant in the fluid dynamics, which is not the case with the values used here. For cool core clusters, we normalise the maximum intensity of the field to 3, 10 and 30~$\mu$G (following values derived by \citealp{EV06}) and we take a value of 1~$\mu$G at the center of non cool core clusters, that corresponds to the same scaling as for the cool core case at 30~$\mu$G (see figure~\ref{fig:B}). 

The coherence lengths of the magnetic field, which are of prime importance for cosmic ray propagation, were calculated directly from the simulated fields, though assuming spherical symmetry for simplicity. We split the cluster in five radial shells  and assume that in each of these shells, the coherence length is roughly constant. The magnetic field at a given distance $r$ from the source can be decomposed into a global and a turbulent components: $B(r)=\langle B\rangle(r) + \delta B(r)$.
We calculate the magnetic turbulence energy spectrum in each radial shell by applying a Fourier transform to the ratio $B/\langle B\rangle=1+\delta B/\langle B\rangle$. We assume in this calculation that the purely turbulent component is defined by an isotropic power law spectrum. This method allows us to get rid of the linear dependance of $\delta B(r)$ over $\langle B\rangle(r)$ and draw a spectrum of the pure turbulence energy. 
The coherence length is then calculated using the following formula (which is valid as long as the turbulent spectrum is steep enough): $\lambda\sim 0.77/k_{\rm min}$, where $k_{\rm min}$ is the wave number at which the energy spectrum reaches its maximal value, setting in the turbulent cascade \cite[]{CLP02}. 
The propagation method implemented in our code enables one to take into account turbulent effects of scale inferior to the grid resolution of the simulation (more details in section~\ref{section:code}). We make use of this advantage by adding by hand a value for the coherence length in a shell of radius smaller than the resolution of the grid ($r=20$~kpc). We set the value of this coherence length to $5$ and 15~kpc for our cool core and non cool core clusters respectively, which are average values according to the studies of \cite{EV06}.   
Table~\ref{table:lcoh} presents our calculated and assumed coherence lengths in each radial shell, for cool core clusters $\lambda_{\rm CC}$ and for non cool core clusters $\lambda_{\rm NCC}$.  \\

\begin{table}[ht]
\begin{center}
  \caption{Magnetic field coherence lengths in kpc}\label{table:lcoh}
  \begin{tabular}{rrrr}
  \tableline\tableline
  $r_{\rm min}$ &  $r_{\rm max}$ & $\lambda_{\rm CC}$ & $\lambda_{\rm NCC}$\\
  \tableline
  0 & 20 & 5 & 15\\
20 & 100 & 36 & 40 \\
100 & 200 & 73 & 80 \\
200 & 800 & 109 & 109\\
800 & 7000 & 145 & 145 \\
  \end{tabular}
  \tablecomments{Coherence lengths in each radial shell (delimited by the distances from the cluster center $r_{\rm min}$ and $r_{\rm max}$), for cool core clusters $\lambda_{\rm CC}$ and for non cool core clusters $\lambda_{\rm NCC}$. Values are in kiloparsecs.}
  \end{center}
\end{table}

\subsection{Possible ultrahigh energy cosmic ray sources in galaxy clusters}\label{subsection:sources}
As mentioned previously, clusters of galaxies are dense regions of the Universe that may harbour various candidate sources of ultrahigh energy cosmic rays. Among those candidates, we are most interested in the framework of this paper, in those located in the inner dense regions of the cluster, where injected particles are better confined and have higher interaction rates, thus some chance of leaving observable signatures. 

Long duration gamma ray bursts \cite[]{Wax95,Vie95,MINN06,MINN08} and some other compact objects like magnetars \cite[]{Aro03} are good candidate sources for ultrahigh energy cosmic rays. These transient sources are expected to be associated with the death of young massive stars which are common in star-forming galaxies (including dwarf and irregular galaxies, see \citealp{LeFloch06, SGL09}). Galaxies with high star formation rate are however not predominant in galaxy clusters, and are particularly rare in the central dense part which is of interest for cosmic ray confinement \cite[]{KK04,Gavazzi06}. \cite{Gavazzi06} report that the star formation rate per unit mass of high luminosity spiral galaxies that are projected within 1 virial radius is about a factor of two lower than at larger clustercentric projected distances. 

Accretion and merger shocks in galaxy clusters were considered by many authors as important cosmic ray accelerators \cite[]{NMA95,KRJ96,KRB97,Miniati00,RK03,IAS05,Inoue07,MIN08}.  
However, high Mach number shocks that are needed to accelerate particles with a relevant spectrum mainly occur at the rim of clusters of galaxies, where the infrared photon and baryon backgrounds are very faint. Therefore, cosmic rays accelerated at such external shocks will experience less interactions in clusters of galaxies than if produced at the center \cite[]{Inoue07}. 
In view of this situation, we will focus in this paper on AGN that are more commonly located in the interesting central region of clusters of galaxies.

What exact part and type of AGN are most susceptible to accelerate particles, and which are more present in galaxy clusters are questions that are beyond the scope of this paper. We might underline however that radio observations indicate a strong presence of radio-loud Faranoff Riley type I (FRI) galaxies in the center of galaxy clusters \cite[]{Best04,Gilmour09}. In cool core clusters especially, as mentioned earlier, these AGN are believed to shut off the cooling flow by mechanical energy dissipation, which would explain the absence of cold gas emission lines in the observations of Chandra and XMM. Powerful Faranoff Riley type II galaxies that are even better candidates for particle acceleration are more likely found in galaxy groups and weak clusters \cite[]{PP88,HL91,Miller02}.

AGN of more quiet types (Seyfert galaxies for example) are also found at an average rate of 1.5 AGN per galaxy cluster \cite[]{Gilmour07, Gilmour09, Martini06}. These studies indicate that these common AGN are predominantly detected in areas of moderate density. For this reason, and mostly because such AGN are much less efficient to accelerate particles at very high energy, these quiet objects are of less interest in our framework.
One might note nevertheless that they may still accelerate protons and nuclei up to high energies, for example in the vicinity of the central black hole or in weak jets. Moreover, it might be possible that some of such AGN were more active in the past, and produced cosmic rays that are still stored in the cluster. These particles could also act as seeds for further acceleration at accretion and merger shocks \cite[]{MIN08}. All these situations introduce many unknown parameters and are not in the scope of this paper.

\subsection{Photonic and baryonic backgrounds}\label{subsection:bg}
We consider in this study the interaction of high energy protons and nuclei with the Cosmic Microwave Background (CMB) and with the infrared background -- the term `infrared' here includes as well ultraviolet and optical backgrounds. We calculate the infrared background by adding the contribution of the diffuse extracluster infrared photon density and the intracluster density created by galaxies inside the cluster.  
We model the diffuse infrared background according to the studies of \cite{KBMH04} and \cite{SMS06}. In this work, we do not include redshift evolution, as its effect is negligible as compared to the uncertainties on all our other parameters, especially as we consider only redshifts of $z\lesssim0.2$.

The intracluster infrared background is modelled as follows. We assume that local ($z\lesssim 0.2$) clusters of galaxies are mostly populated by elliptical galaxies that moderately enrich the cluster with infrared photons. We thus scale an average elliptical galaxy SED (Spectral Energy Distribution) with the galaxy density in our simulation cubes, convoluted by a $1/r^2$ function in order to account for the flux diminution with distance. The SED was kindly provided by Herv\'e Dole and details of their modelling can be found in \cite{LPD05} and in references therein. The colour contours of the projected infrared photon density are drawn in the middle panel of Fig.~\ref{fig:3slices}.

It is important to stress again that the galaxies in galaxy clusters are ellipticals in majority. According to the analysis of \cite{Holden07} and \cite{Vanderwel07}, the fraction of ellipticals among galaxies of mass $M>4\times 10^{10} \rm M_\odot$ in clusters at redshift $z<0.2$ is higher than 90\%. These galaxies have SED that highly differ from those of ultra-luminous infrared galaxies (ULIRG) that were used by \cite{demarco06} to evaluate their cluster infrared background. 

Another important process for cosmic rays propagating in the intracluster medium, especially at energies $E\lesssim10^{18}$~eV, is their interaction with the cluster gas. The baryonic background is evaluated using the simulation outputs from \cite{DT08}. In order to benefit from the high resolution of the simulation in the densest region of the cluster, we do not use a smoothed out three dimensional mapping of the density grid, but the mean radial profile instead, with high resolution at the center. We checked that the density distribution is actually smooth enough over the whole simulation box, so that the loss of small features while averaging over radial bins does not affect our results. 

\begin{figure}[htb]
\begin{center}
\includegraphics[width=0.4\textwidth]{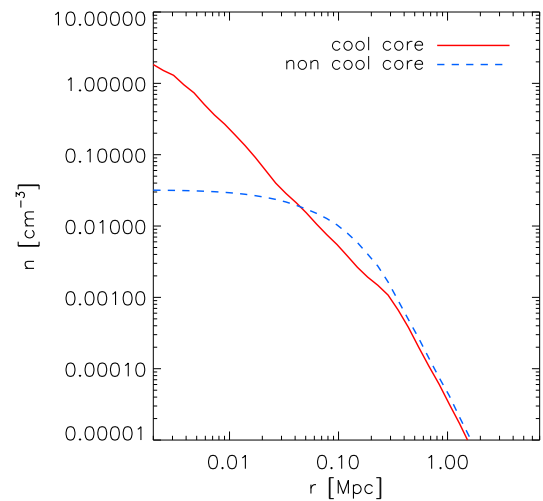} 
\caption{Baryonic density profiles from the simulations of \cite{DT08}. The red solid line represents the mean baryonic density in radial bins for the cool core cluster of galaxies. The blue dashed line is for the non cool core cluster.}  \label{fig:rho}
\end{center}
\end{figure}

\section{Propagation of protons and nuclei: numerical techniques}\label{section:code}

\begin{figure*}[htb]
\begin{center}
\includegraphics[width=0.33\textwidth]{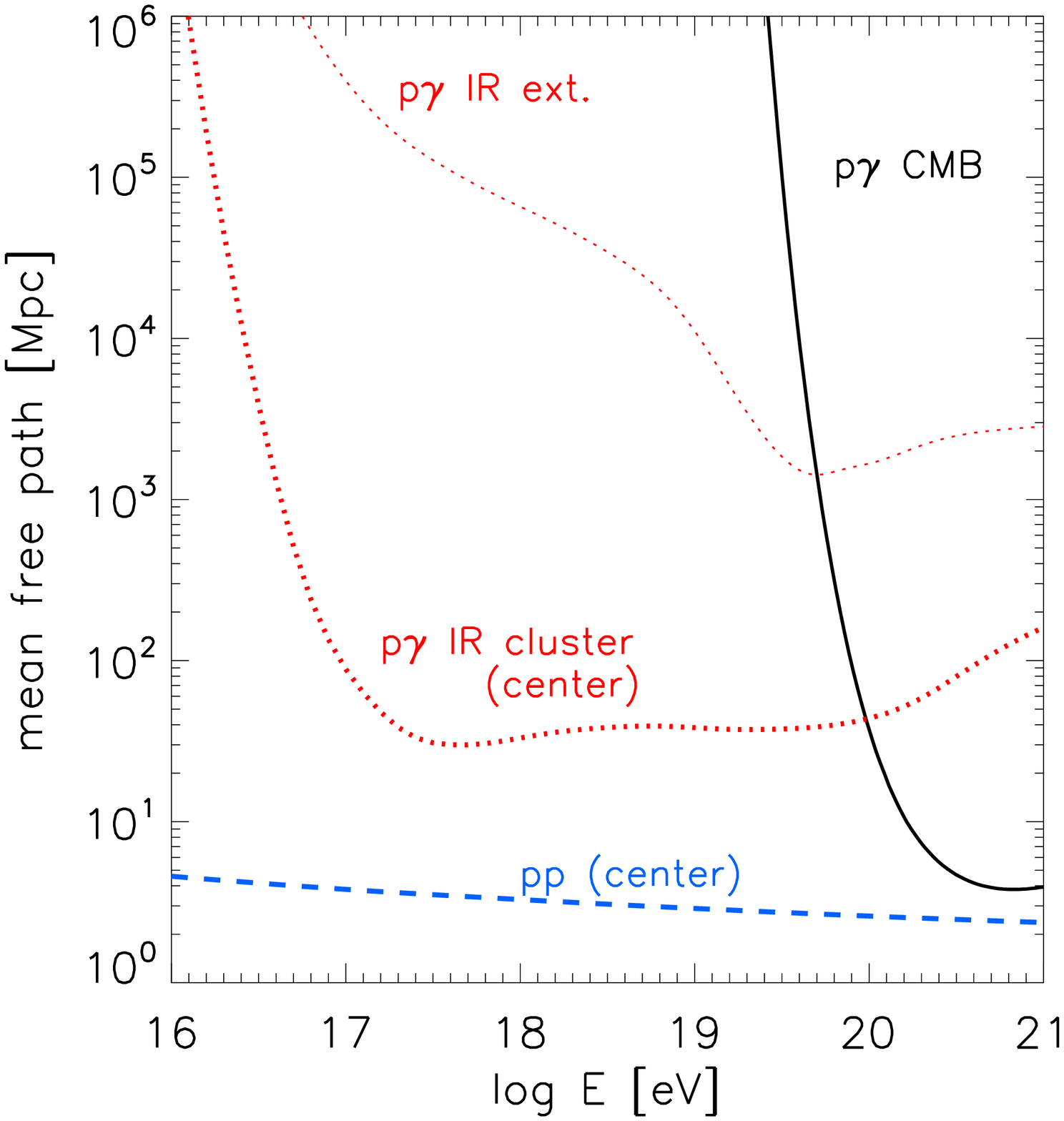} 
\includegraphics[width=0.33\textwidth]{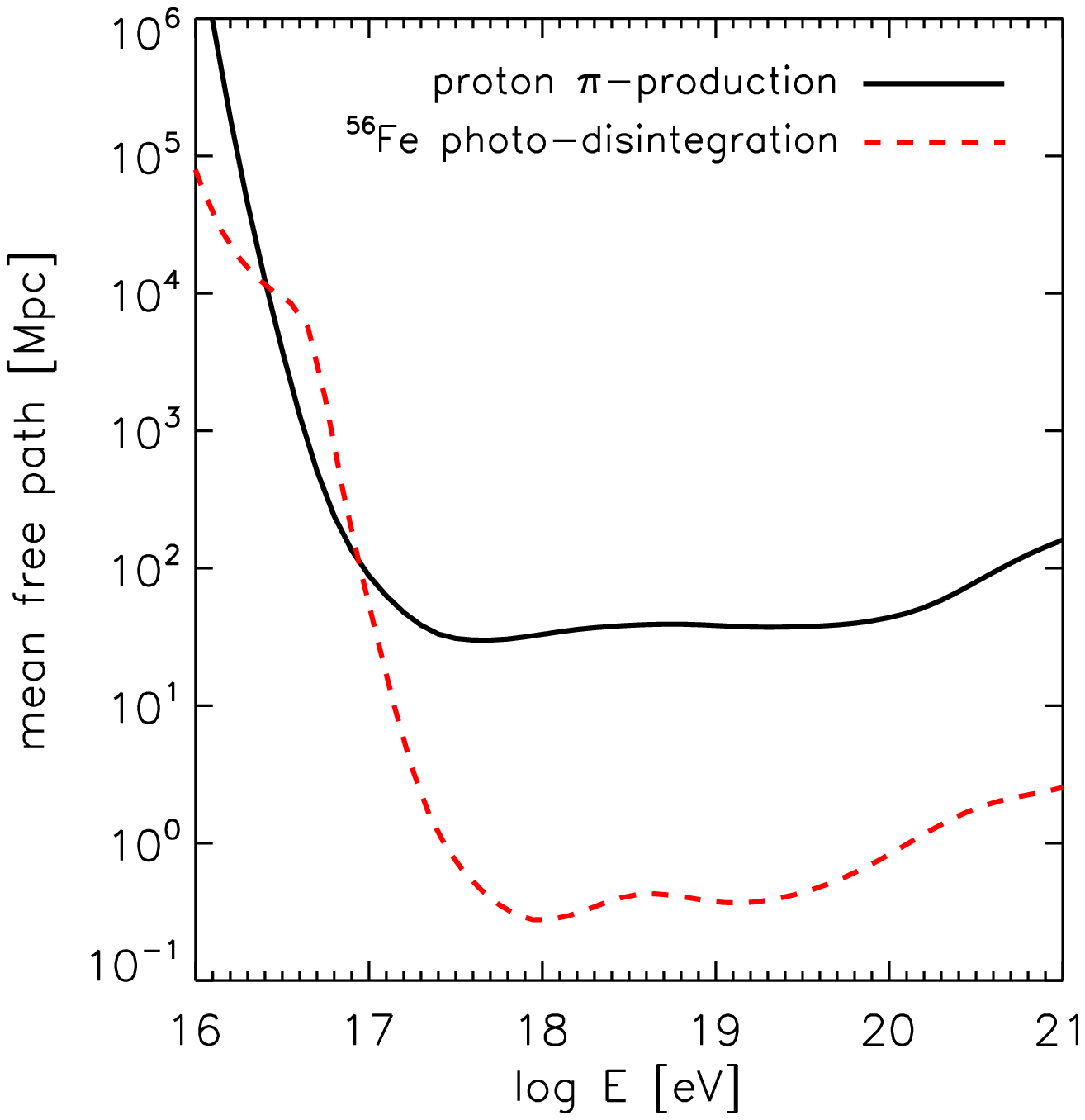} 
\includegraphics[width=0.33\textwidth]{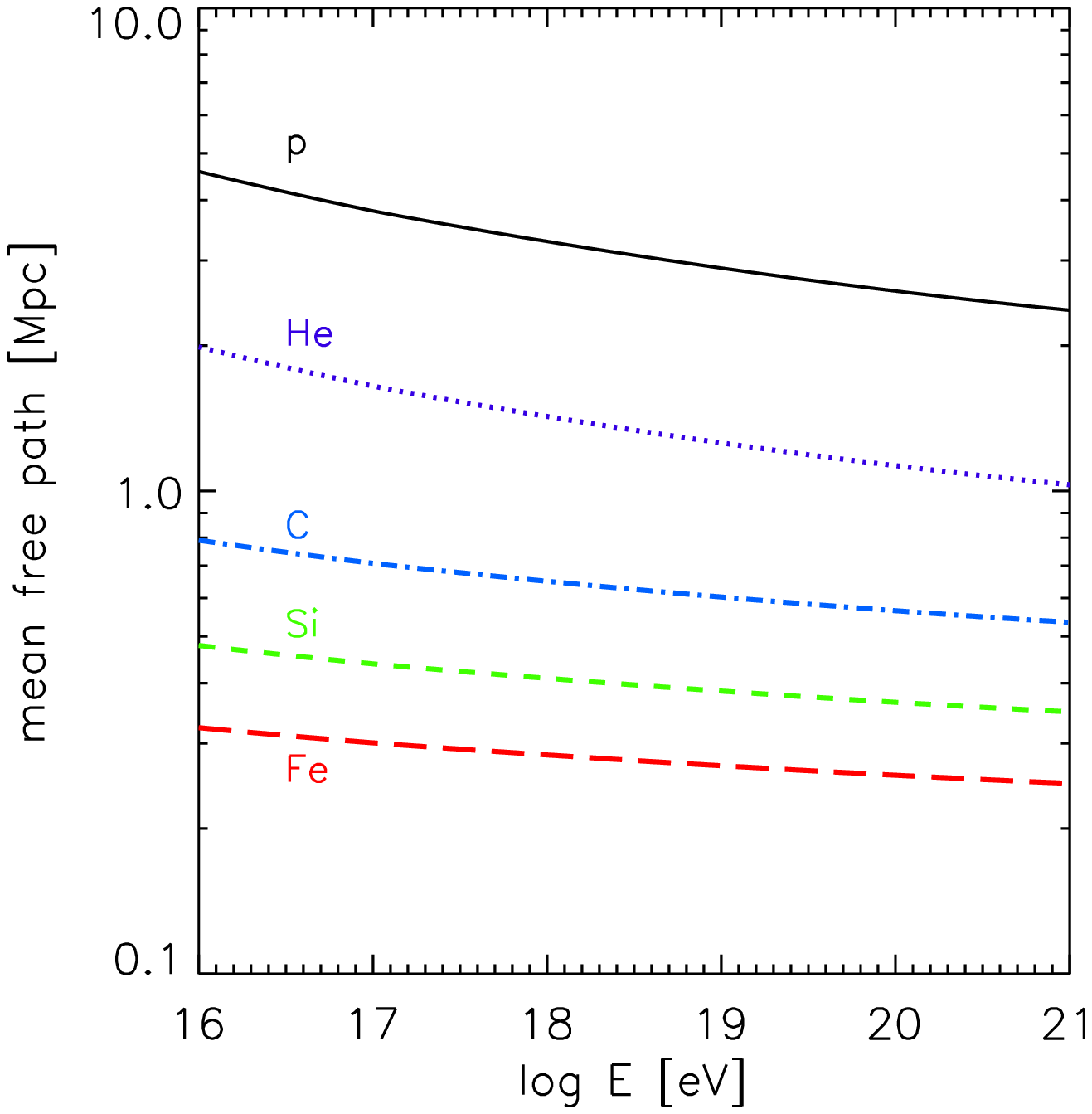} 
\caption{Mean free paths as a function of cosmic ray energy for various interactions with various backgrounds, all calculated in the inner 20~kpc region of the cool core cluster. {\it Left panel: } values for a propagating proton. Pion photo-production on the CMB (black solid line), on the diffuse extragalactic infrared background (red thin dotted line), on the cluster infrared background (red thick dotted line) and hadronic interactions on the baryonic gas (blue dashed line).  {\it Middle panel:} comparison between pion photo-production process (black solid) and photo-disintegration for iron (red dashed) for interactions with the cluster infrared background. {\it Right panel: } variation of the hadronic interaction mean free paths for (from top to bottom) proton, helium, carbon, silicon and iron. The cross sections used in these calculations were provided by EPOS.
 }  \label{fig:MFP}
\end{center}
\end{figure*}

We describe in this section some aspects of our ultrahigh energy nuclei propagation code, that combines a fast and accurate semi-analytical trajectory integration method in the magnetic field, and complete Monte Carlo calculations of photonic and baryonic energy losses for primary and secondary nuclei. \\

The transport scheme in the magnetic field was adapted from \cite{KL08a}. In this method, particles propagate by crossing spheres of diameter the local coherence length of the magnetic field. The exiting time and deflection angle of the particle when leaving the sphere are randomly computed following analytical distribution functions depending on the field strength, coherence length and Larmor radius of the particle. These functions include information on the small scale turbulence component via the diffusion coefficient that we implemented according to the results of \cite{CLP02}. The advantages of this method are twofold: the trajectory integration is very fast as compared to a direct integration, and second, it allows to enlarge artificially the range of scales on which the magnetic field is distributed, and thus to account for turbulence scales below the simulation grid resolution. More details on this scheme can be found in the appendix of \cite{KL08a}.\\

The interactions of protons and nuclei with CMB, infrared, optical and ultraviolet photon backgrounds were mostly modelled according to the Monte Carlo methods of \cite{Allard05, Allard06}. These interactions produce features in the propagated ultrahigh energy cosmic ray spectrum such as the ``GZK cutoff'' \cite[]{G66,ZK66} and their decay products generate the cosmogenic neutrino flux \cite[]{BZ69}. During the propagation in the intracluster medium, the same interactions are also expected to take place, at a rate depending on the ambient photon density and the confinement time in the densest regions, that were described in the previous section. These interactions can also possibly produce spectral and composition features in the cosmic rays as well as secondary neutrinos and photons.

In our calculations, electron and positron pair production from the interactions with the various photon backgrounds (also known as the Bethe-Heitler process) are treated as a continuous energy loss process. We use the cross sections, inelasticities, as well as the mass and charge scaling of the attenuation length for nuclei that are provided by \cite{R96}.  

If the energy of the background photons exceeds $\sim$145~MeV in the nucleon rest frame, protons and neutrons can interact through the pion photoproduction process. This latter process is taken into account in our code using Monte Carlo calculations based on the outputs of event generator SOPHIA \cite[]{M00}. The use of SOPHIA allows us to treat accurately the various interaction channels (direct pion production, resonances, multi-pion production) and their branching ratios (see \citealp{R96}).

The interactions experienced by nuclei with photon backgrounds differ from those of protons. In addition to the pair production losses that result in a decrease of the Lorentz factor and the rigidity of the nucleus, one must consider the photo-disintegration  processes that lead to the ejection of one or several nucleons from the nucleus. 

Different photo-disintegration processes become dominant in the total interaction cross section at different energies \cite[]{PSB76}.  The lowest energy disintegration process is the Giant Dipole Resonance (GDR) which results in the emission of one or two nucleons and $\alpha$ particles. The GDR process is the most relevant as it has the highest cross section and the lowest thresholds, between 10 and 20 MeV for all nuclei. For nuclei with mass $A \geq 9$, we use the theoretically calculated GDR cross sections presented by \cite{Khan05},
which take into account all the individual reaction channels and are in better agreement with data than previous treatments.  For nuclei with $A < 9$, we use the phenomenological fits to the data provided by \cite{R96}.  Around 30 MeV in the nucleus rest frame, and up to the photopion production threshold, the quasi-deuteron (QD) process becomes comparable to the GDR and dominates the total cross section at higher energies.
The photopion production (or baryonic resonances, BR) of nuclei becomes relevant above 150 MeV in the nucleus rest frame (e.g., $\sim5\times 10^{21}$ eV in the lab frame for iron nuclei interacting with the CMB). We use the parameterisation given by \cite{R96} in which the cross section in this energy range is proportional to the mass of the nucleus. For more details on the numerical method, one might refer to \cite{Allard05}.\\

In addition to interactions with soft photon backgrounds, protons and nuclei can experience hadronic interactions with the baryonic background in the dense regions of the intracluster medium. We implement this process using the hadronic interaction model EPOS \cite[]{WLP06}. We assume that the baryonic background is essentially made of protons at rest, which is a good approximation in view of the extremely high Lorentz factors of the projectile nuclei. Our choice of using the new interaction model EPOS was motivated by the fact that it is the only model for very high energy cosmic ray interactions which is actually also used for particle physics analysis. It has indeed been tested thoroughly against most of the existing hadron-hadron and hadron-nucleus data. As a consequence, EPOS has a very good description of secondary products, which are of interest to our work. In EPOS, the fragmentation of heavy nuclei is not treated: spectator nucleons are set as free particles and the building up of resulting nuclei has to be done as post-processing.  
We make use for this purpose of the fragmentation model of \cite{CH81}, as implemented in the air shower simulation code CONEX \cite[]{Bergmann07}.\\

Figure~\ref{fig:MFP} shows the mean free paths for the interactions described above, as a function of the cosmic ray energy, for the cool core cluster. Values for the non cool core case can be derived from these plots by simple scaling laws.  

The first plot presents the mean free paths for interactions experienced by protons, namely pion production and hadronic interactions -- mean free paths for pair production are not represented. We draw separately the contribution of the CMB photon background (black solid line), the diffuse extragalactic infrared background (thin dotted red line) and in thick dotted lines the infrared background of the cluster itself, in the inner region (with distance from the center $r\lesssim 20$~kpc). It clearly appears that the interactions with the CMB photons dominate at extreme energies ($E\gtrsim10^{20}$~eV). One can also notice that the infrared background at the center of the cluster is highly enhanced as compared to the diffuse extragalactic background, resulting in a shortening of the mean free paths around $E\sim 10^{18}$~eV of more than 4 orders of magnitudes. One should however note that the extragalactic background becomes dominant for energies $E\lesssim 10^{19}$~eV at $\sim1$~Mpc of distance from the center of the cluster. Obviously, the contribution of hadronic interactions will be predominant in the center of the cluster, at all energies. Mean free paths for proton-proton interactions at different distances from the center can be easily extrapolated from this plot by scaling with the baryonic density presented in Fig.~\ref{fig:rho}. Due to the rapid decrease of the density with the distance from the center of the cluster, hadronic interactions become quickly negligible and are completely overcome by photonic interactions at 0.1~Mpc. 

The figure in the middle panel compares the mean free paths for interactions of protons (black solid line) and iron (red dashed line) with the cluster infrared background, as a function of their energy. The distances are calculated again in the inner part of the galaxy cluster ($r\lesssim 20$~kpc). The mean free path for photo-disintegration of iron -- that cumulates all GDR, QD and BR processes -- are shorter by more than two orders of magnitude than for the pion photo-production with protons. This implies that heavy nuclei can be strongly depleted by this process at the center of the cluster, but not necessarily that secondary photons and neutrinos will be emitted. Indeed, the only process here that produces pions is the baryonic resonance (BR) which takes place only at extremely high energy. In comparison, pion production from proton-photon interactions can happen at all energies above $E\sim10^{17}$~eV. The depleted nuclei can still produce secondary protons that might interact through this channel. It appears that accurate calculations are definitely necessary to investigate the possible signatures of these different processes.

The last panel of Fig.~\ref{fig:MFP} presents the variation of the hadronic interaction mean free paths for a set of nuclei, as a function of their energy, in the inner 20~kpc of the cluster. The cross sections used in these calculations were provided by EPOS. It appears that at high energy, the mean free paths for nuclei can be derived from the proton-proton interaction case by a simple scaling law, but there is a slight discrepancy at lower energy. As expected, heavy nuclei have higher probability (of order $A^{2/3}$ times more, where $A$ is the nuclei mass number) to interact with the cluster gas, leading to the production of a wide bunch of secondary particles.

\section{Results and discussion}\label{section:results}

In this section, we present the results of our simulations for various normalizations of the magnetic field, in the case of cool core and non cool core clusters of galaxies. We position the source of ultrahigh energy cosmic rays at the center of the cluster, or at a slightly shifted position, and inject either a 100\% proton composition, or a mixed composition as in \cite{Allard06}, based on Galactic cosmic ray abundances. The injection spectral index is chosen to be $2.3$ for all our plots (if not indicated otherwise) and a maximum injection energy of $E_{{\rm max},Z}=Z\times10^{20.5}~$eV is assumed for a nucleus of charge number $Z$ (an exponential cut-off is taken above). Most of our results are arbitrarily normalised to unity at $E=10^{19}$~eV. When a comparison with observational data or limits are needed, we normalise our fluxes by choosing the cosmic ray luminosity of the central AGN to be $L_{\rm cr} = 10^{45}$~\ergpers{} with a minimum injection energy $E_{\rm min}=1$~GeV. This integrated luminosity indeed allows to match the ultrahigh energy cosmic ray flux assuming a source density of $n_{\rm s}=10^{-5}$~Mpc$^{-3}$, for this value of the spectral index. Moreover such a luminosity is reasonable for this type of object. 

\begin{figure}[htb]
\begin{flushleft}
\includegraphics[width=0.45\textwidth]{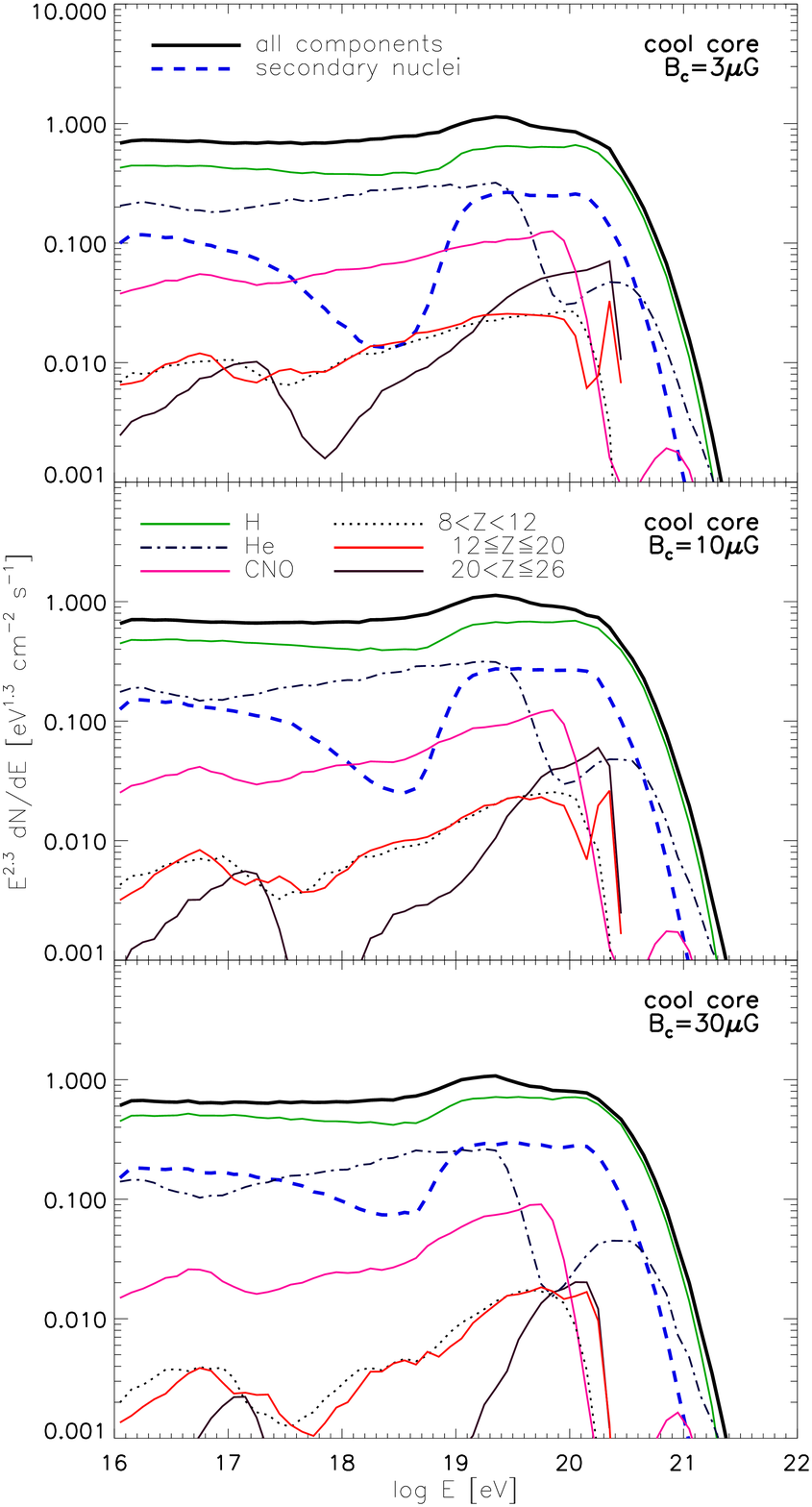} 
\caption{Cosmic ray energy spectra for central injection and for our cool core cluster with magnetic field at the center: $B_{\rm c} = 3~\mu$G, $10~\mu$G and $30~\mu$G from top to bottom. The black solid line presents the total cosmic ray flux and the blue dashed line the flux of secondary nuclei produced during the propagation. We also show the contribution of the different chemical species to the total flux as indicated in the second panel: protons (green), Helium (black dot-dashed), Carbon, Nitrogen and Oxygen (pink), all species with charge $Z$ comprised between 12 and 20 (red) and heavy nuclei with charge $20<Z<26$ (brown). These spectra are normalised to unity at $E=10^{19}$~eV.  }  \label{fig:nuclei_spectra_cc}
\end{flushleft}
\end{figure}

\begin{figure}[htb]
\begin{flushleft}
\includegraphics[width=0.45\textwidth]{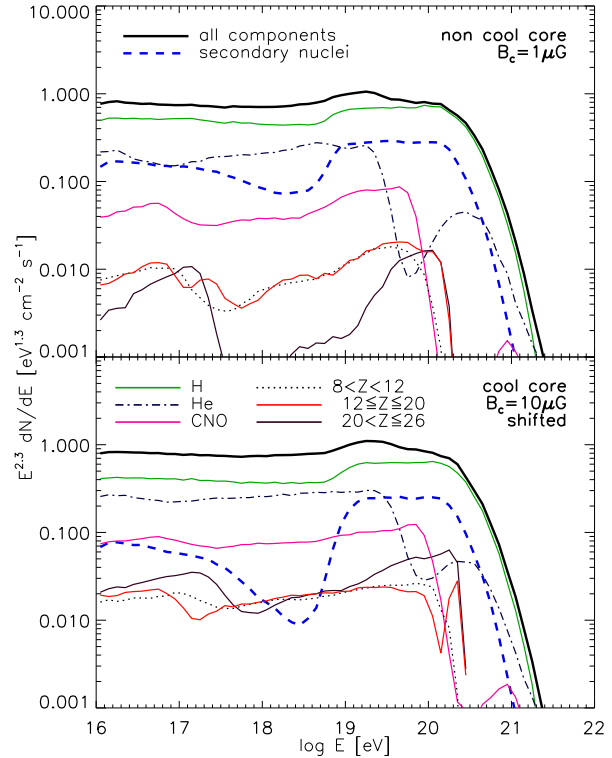} 
\caption{Same as Fig.~\ref{fig:nuclei_spectra_cc} but for our non cool core cluster with $B_{\rm c}=1~\mu$G  and central injection ({\it top}), and for our cool core cluster with $B_{\rm c} = 10~\mu$G and the source positioned at 100~kpc from the center ({\it bottom}).
}  \label{fig:nuclei_spectra_other}
\end{flushleft}
\end{figure}

\begin{figure*}[htb]
\begin{flushleft}
\includegraphics[width=\textwidth]{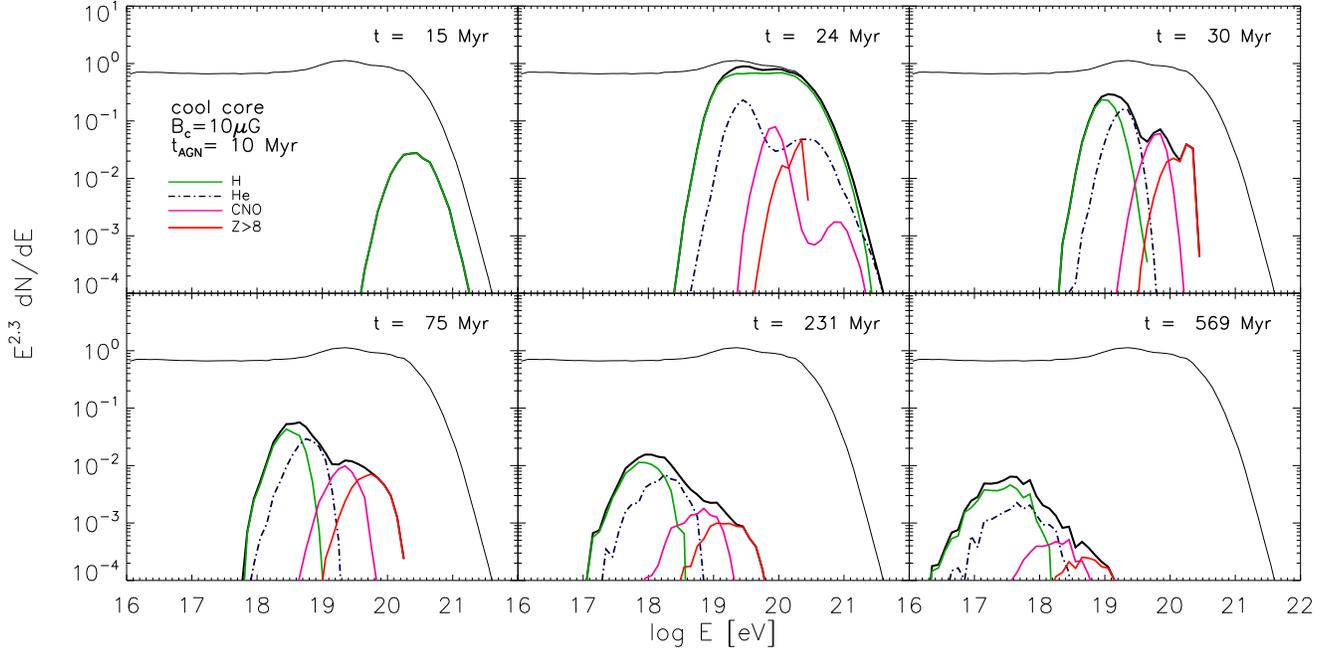} 
\caption{Evolution of the cosmic ray spectrum in time, assuming a lifetime of $t_{\rm AGN}=10$~Myr for the central AGN, for the case of a cool core cluster of central magnetic field $B_{\rm c}=10~\mu$G. Each panel presents the spectrum at the time indicated at the top-right hand corner. The injection from the source (AGN) is assumed to begin at $t=0$. The contribution of the different chemical components are shown in the same color code as for Fig.~\ref{fig:nuclei_spectra_cc}. The thick black line is the total spectrum and the thin black line indicates the total flux obtained for an infinite AGN lifetime and an integration of the flux over a Hubble time (stationary regime), as in Fig.~\ref{fig:nuclei_spectra_cc}. The spectra are normalised to the value of the stationary flux obtained at $E=10^{19}$~eV.
 }  \label{fig:tagn10}
\end{flushleft}
\end{figure*}

\begin{figure*}[htb]
\begin{flushleft}
\includegraphics[width=\textwidth]{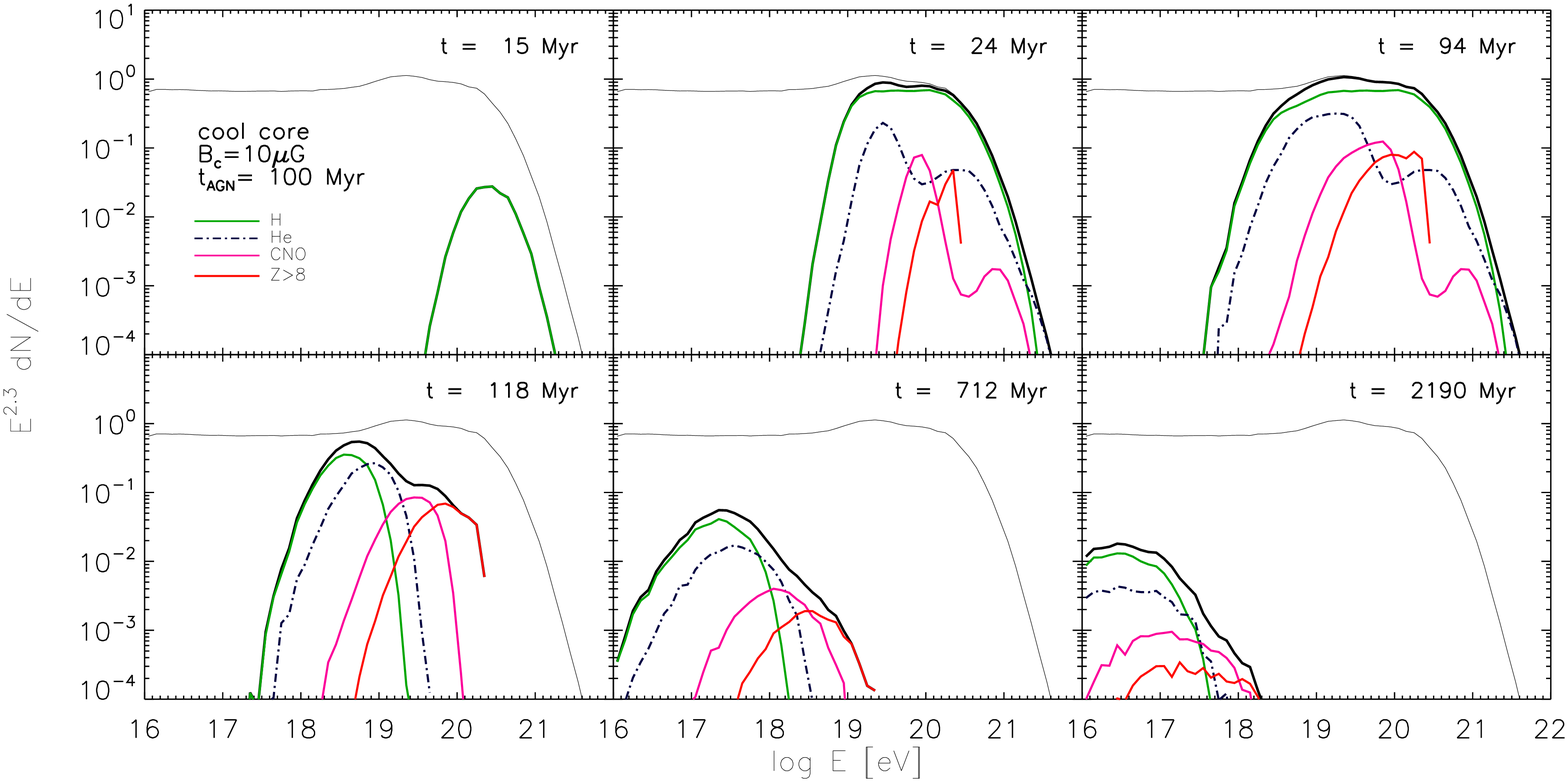} 
\caption{Same as Fig.~\ref{fig:tagn10}, but with an AGN lifetime of $t_{\rm AGN}=100$~Myr. 
 }  \label{fig:tagn100}
\end{flushleft}
\end{figure*}

\subsection{Cosmic ray spectra}\label{subsection:crspec}

One key point of this paper is to investigate the resulting composition of ultrahigh energy cosmic rays, when a mixed chemical composition is injected at the center -- or at a slightly shifted position -- of a cluster of galaxies.

Figure~\ref{fig:nuclei_spectra_cc} shows our resulting cosmic ray spectra for our cool core cluster of galaxies for various normalizations of the magnetic fields (at the center: $B_{\rm c} = 3~\mu$G, $10~\mu$G and $30~\mu$G from top to bottom).  The black solid line presents the total cosmic ray flux and the blue dashed line the flux of secondary nuclei produced during the propagation. We also present the contribution of different chemical species to the total flux as indicated in the second panel: protons (green), Helium (black dot-dashed), Carbon, Nitrogen and Oxygen (pink), all species with charge $Z$ comprised between 12 and 20 (red) and heavy nuclei with charge $20<Z<26$ (brown). These three panels assume that the injection of particles occurs at the very center of the cluster, where the background densities are the highest. The limited lifetime of AGN is not considered in these plots: we assume that the source emits particles continuously and that the stationary regime has been reached. 

In Figure~\ref{fig:nuclei_spectra_cc} one can observe features in the spectra of the different nuclei, that increase in amplitude with the mass of the nuclei and the normalization of the magnetic field. At the lowest energies the flux of the heaviest nuclei is greatly suppressed due to hadronic interactions. In the same energy range, light and intermediate nuclei suffer less interactions due to their larger mean free paths ($l_{Ap}\propto A^{-2/3}$) and rigidities (as the confinement time $\tau_{\rm conf}\propto (E/Z)^{-1/3}$). As the baryonic density rapidly decreases with the distance to the center and the confinement time with the energy, hadronic interactions become less efficient to disintegrate nuclei above $E\sim 10^{17}$~eV. Above this energy however, heavy nuclei start to interact with optical and near-infrared photons. The photon density decreases more slowly than the baryonic density; photo-disintegration is then very efficient to deplete the heavy nuclei component especially for the largest magnetic field normalizations. For lighter nuclei, the energy threshold for photo-interactions is lower (more or less proportional to the mass of the nuclei $A$) but the mean free path is larger (more or less proportional to $A^{-1}$), therefore the effect of photodisintegration on light and intermediate nuclei is weaker. 

The production of secondary protons and their relative abundance at low energies depend on the interactions of nuclei. One can see as expected, that the relative abundance of secondary protons is higher for the largest magnetic field normalization due to the better confinement in  the vicinity of the cluster center. Between $E\sim10^{17.5}$ and $10^{18.5}$ eV the decrease of the secondary proton abundance results from the deconfinement of nuclei and can be interpreted as a counterpart of the recovery of the different nuclei components at higher energies (especially visible for the heaviest nuclei). As mentioned above, this decrease is slower for the largest magnetic field case. Above $E\sim 4\times10^{18}$ eV, the secondary proton fraction increases again due to the interaction of nuclei with CMB photons, starting at the energy threshold $\sim A\times 4\times10^{18}$ eV. Nuclei are completely disintegrated by CMB photons and turn into secondary nucleons of lower energies, resulting in an expected $\sim 100\%$ proton composition above $E\sim3\times10^{20}$ eV. Let us note nevertheless that these interactions (unlike those at lower energies) are not due to the cluster environment, as nuclei are barely confined by the cluster magnetic field at these energies; they would take place in a similar way in the intergalactic medium (see for instance \citealp{Allard08}). 

These figures demonstrate that with a high (but realistic) magnetic field of $B_{\rm c}\sim 30~\mu$G at the center of a cool core cluster, the heaviest nuclei hardly survive, and escape the structures only for energies around $E\sim 10^{20}~$eV. For lower magnetic fields, though, a reasonable amount of heavy nuclei can still survive the propagation inside clusters of galaxies. 
One can also note that no heavy nuclei survive at the highest energies ($E\gtrsim10^{20.5}$) due to photo-disintegration on the CMB photons, meaning that the composition becomes 100\% protons in this region. The small fractions of light fragments ($Z<6$) present at the highest energies after propagation inside the cluster will be rapidly photo-disintegrated after a few Mpc in the extragalactic medium.\\

Figure~\ref{fig:nuclei_spectra_other} presents the cases of a non cool core cluster with central magnetic field $B_{\rm c}=1~\mu$G with the source located at the center (top panel), and of a cool core cluster with $B_{\rm c} = 10~\mu$G and the source positioned at 100~kpc from the center (bottom panel). The upper panel should be compared to the lowest panel of Fig.~\ref{fig:nuclei_spectra_cc}, as the scaling coefficient is the same in both cases, meaning that the magnetic field profiles differ here only in the core of the cluster (see Fig.~\ref{fig:B}). The higher magnetic field and the enhanced baryonic density in the cool core case (see Fig.~\ref{fig:B} and \ref{fig:rho}) definitely play a role at lowest energies ($E\lesssim 10^{17.5}$~eV): the confinement in the core is more efficient and heavy nuclei are more depleted by hadronic interactions than in the non cool core case. The smaller coherence length in the center of the cool core cluster (see table~\ref{table:lcoh}) also contributes to increasing the confinement time in the Kolmogorov diffusion regime, as $\tau_{\rm conf}\propto Z^{1/3}E^{-1/3}B^{1/3}\lambda^{-2/3}$, with $\lambda$ the coherence length of the magnetic field. 

The fact that the two first panels of Fig.~\ref{fig:nuclei_spectra_cc} appear more optimistic for the survival of heavy nuclei than for the non cool core case in spite of a larger magnetic field at the center is due to our scaling of the magnetic field. The field being peaked at the center for the cool core case, an overall normalization of $B$ to obtain a required value of $B_{\rm c}$ at the center lowers also the intensity of the field at the rim of the core. We thus get a magnetic field that becomes quickly very faint as we move from the center, and in particular, it will be fainter than for the case of a cool core cluster normalised at $B_{\rm c}=1~\mu$G. These first two plots should actually be compared to cool core clusters normalised at $B_{\rm c}=0.3~\mu$G and $0.1~\mu$G respectively.

It is very plausible that the acceleration sites of ultrahigh energy cosmic rays are not at the very center of the cluster of galaxies, but shifted of some hundreds of kiloparsecs. This would be especially the case of particles are accelerated at hot spots \cite{RB93} or  in lobes of radio galaxies (e.g., \citealp{ORT09} and references therein), which form at a distance of some hundreds of kiloparsecs from the central black hole \cite[]{BP84}. The lower panel of Fig.~\ref{fig:nuclei_spectra_other} presents the resulting spectrum obtained if the injection of ultrahigh energy cosmic rays happens at 100~kpc of the center of the cluster. In a cool core cluster, the baryonic density falls quite steeply with the distance from the center: at 100~kpc, the average density is already two magnitudes lower than at the center. This explains why nuclei survive much better in this case, especially at low energy. The features in the heavy nuclei component we described qualitatively above are still present but with a much lower amplitude. The initial composition is in this case less modified by the propagation in the cluster environment. The position of the source will thus have a strong impact on the resulting composition of ultrahigh energy particles and an ample depletion of the heavy component can only be expected if the source is in the immediate vicinity of the cluster center. \\

The effects of propagation in the extragalactic medium are not included in these results but they should not affect importantly the composition for nearby sources (at less than $100-200$~Mpc). Indeed, the interactions with the CMB has already depleted the nuclei inside the cluster and hence further interactions will be negligible at ultrahigh energies. Some influence of the far infrared background might play a slight role for particles of energy $E\gtrsim 3\times10^{19}$~eV, diminishing the flux of heavy nuclei, but this should also be a minor effect compared to interactions experienced in the cluster. Hence at the highest energies, the composition should not change, we however expect a change in the shape of the spectrum due to the GZK cut-off for the remaining protons.
Magnetic horizon effects due to the diffusion of low energy cosmic rays in the extragalactic magnetic field should introduce a cut-off for energies $E\lesssim 10^{17.5}$~eV if the source is located at more than $\sim$10~Mpc, for some magnetic field configuration and strengths (see \citealp{L05,KL08a,Globus08}). The extragalactic magnetic field structure and intensity being very poorly known, we prefer not to introduce a large amount of new parameters, and do not compute this effect. 

In order to compare the produced cosmic ray fluxes with the available observed datasets of  AGASA, HiRes and the Pierre Auger Observatory, we computed the diffuse flux assuming the same mass for all clusters, and a number density of $n_{\rm s}=10^{-5}$~Mpc$^{-3}$. Note that massive clusters with $M \gtrsim {10}^{15}~{M}_{\odot}$ are rare objects of number density $2 \times {10}^{-6}~{\rm Mpc}^{-3}$, but the density can be a few times~${10}^{-5}~{\rm Mpc}^{-3}$ when one includes relatively smaller clusters of $M \gtrsim 5 \times {10}^{14}~{M}_{\odot}$ \cite[]{Jenkins01}. Assuming that some fraction only of those clusters host FRI type objects, the density that is of interest to us falls around $\sim10^{-5}~$Mpc$^{-3}$. We propagate the cosmic rays exiting the cluster environment through the extragalactic medium (including all the relevant energy loss processes), and find that the diffuse cosmic ray flux fits very well the observed fluxes for the range of parameters we chose (namely $L_{\rm cr}=10^{45}$~\ergpers{}, which is an upper limit for the luminosity of Seyfert and FRI galaxies, and a 2.3 spectral index). This implies that higher luminosity sources or higher galaxy cluster densities would result in an overproduction of the ultrahigh energy cosmic ray flux. The normalization of the spectrum indeed scales as $L_{\rm cr}\times n_{\rm s}$. One may also note that a harder injection spectral index will have the same effect of overproducing particles as compared to data. \\

We have considered up to now spectra calculated in the case of a permanent regime, assuming an infinite AGN lifetime. It is yet of common knowledge that AGN remain active only during a limited time. Studies indicate that the typical lifetime of radio sources is of order $\sim 10^{7-8}$~yrs for FRI type galaxies and somewhat shorter for FRII \citep{Parma02,Bird08}. Such a short injection time can have a considerable effect on particle spectra, as one can notice that these times are much shorter than the confinement times of some low energy nuclei. 

Figures~\ref{fig:tagn10} and \ref{fig:tagn100} present the evolution of the cosmic ray spectra in time, assuming a limited AGN lifetime of $t_{\rm AGN} = 10$~Myr and 100~Myr respectively. The cosmic ray afterglow observed after the extinction of the source is due to the confinement times of different species at different energies and to their variance around their mean value. While time goes, we observe the progressive apparition of low energy particles and of heavier nuclei. The variance $\sigma_{\rm conf}$ around the confinement time $t_{\rm conf}$ is globally proportional to the latter, meaning that high energy light nuclei have a small variance. For this reason, high energy protons and Helium quickly disappear as the source dies, while heavy nuclei with larger confinement time and variance remain present a much longer time. This leads interestingly to a heavy composition at the highest energies for times greater than $\sim t_{\rm AGN} + t_{\rm esc}$, where $t_{\rm esc}$ is the escaping time of protons propagating rectilinearly from the cluster. After $\sim 10\, t_{\rm AGN}$, the flux is considerably diminished at all energies. 

Such effects could be detectable if a few nearby clusters of galaxies with an extinguished AGN contribute significantly to the observed diffuse flux of ultrahigh energy cosmic rays. It is also interesting to notice that such effects should be present in the case of any other type of non stationary sources embedded in magnetised media and injecting a mixed composition. One might also relate these time dependent fluxes to the absence of powerful sources in the arrival directions of the observed highest energy events: magnetised clusters of galaxies hosting an extinguished AGN can be emitting cosmic ray afterglows, and contribute to the overall observed spectrum, provided that their density and source luminosity are high enough. 

These time dependent effects will occur according to the duration of the duty cycle of AGN. This value will set the average cosmic ray emission state at which one expects to observe a cluster of galaxy hosting an AGN. We may also notice that in the presence of more than one AGN at the center of the cluster, there might be a spread in the effective injection duration, which can mimic a permanent regime. In such a case, the obtained fluxes are those calculated in Fig.~\ref{fig:nuclei_spectra_cc} and \ref{fig:nuclei_spectra_other}.

\subsection{Secondary neutrinos}

Another signature of the propagation of ultrahigh energy nuclei in clusters of galaxies can be found in the produced secondary neutrinos. One should note however, following the previous discussion, that if the ratio between the durations of inactivity and activity phases of some AGN in clusters of galaxies is important, we might not observe neutrinos and photons simultaneously with ultrahigh energy cosmic rays. 

\begin{figure}[tb]
\begin{flushleft}
\includegraphics[width=0.45\textwidth]{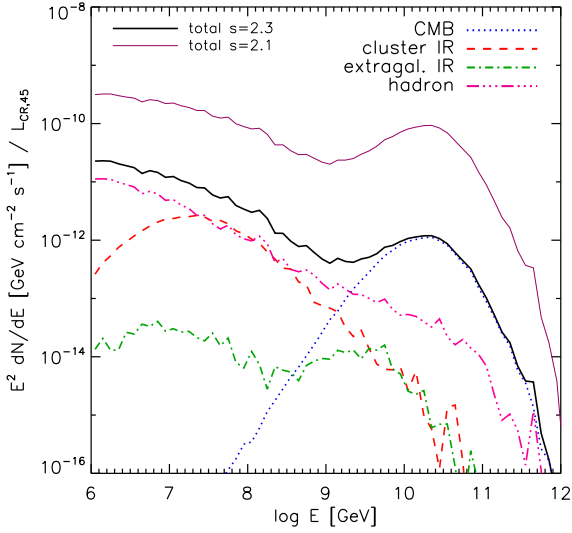} 
\caption{Neutrino flux for a cool core cluster with central magnetic field $B_{\rm c}=10~\mu$G and a source located at the center, injecting a mixed galactic composition with $L_{\rm cr}=10^{45}$~erg~s$^{-1}$, at a distance of $100$~Mpc, for spectral index 2.3 (thick black) and 2.1 (thin violet). The contribution of the different backgrounds are indicated for spectral index 2.3: interactions with the CMB photons (blue dotted), with the infrared photons produced by galaxies of the cluster (red dashed), with the diffuse extragalactic background (green dash dotted), and with the baryonic background (pink dash dot dotted).
 }  \label{fig:neut}
\end{flushleft}
\end{figure}

Figure~\ref{fig:neut} presents our calculated neutrino flux from a single cool core cluster of galaxies located at 100~Mpc, and embedding a central source of cosmic ray luminosity $L_{\rm cr}=10^{45}$~\ergpers{} that injects a mixed composition with spectral index 2.3 with maximum energy $E_{\rm max}=10^{20.5}$~eV. We do not inject particles with energy less than $E_{\rm min,simu}=10^{16}$~eV and stop propagating nuclei when they reach this threshold: consistently, we do not present the flux of neutrinos below 1~PeV. These low energy neutrinos would in any case be rapidly overwhelmed by atmospheric neutrinos and make any detection (especially for the diffuse flux) more difficult. This plot enables one to measure the contribution of the various backgrounds on the final neutrino flux. As expected, the hadronic interactions play a major role at all energies, and the cluster infrared background contributes around $E\sim 10^{17.5-19}$~eV, leading to the production of neutrinos of energy $10^{7-8}$~GeV. 

\begin{figure}[tb]
\begin{flushleft}
\includegraphics[width=0.45\textwidth]{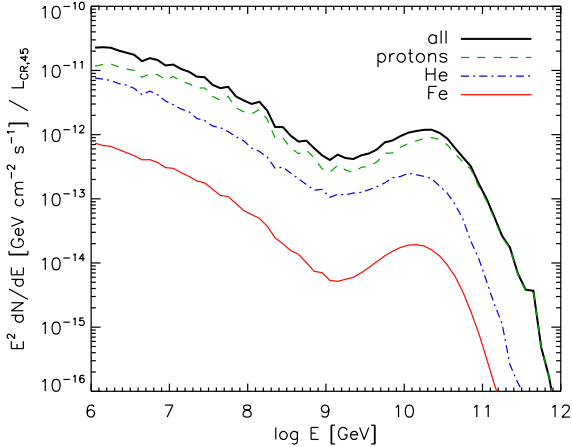} 
\caption{Neutrino flux for a cool core cluster with central magnetic field $B_{\rm c}=10~\mu$G and a source located at the center, injecting a mixed galactic composition with $L_{\rm cr}=10^{45}$~erg~s$^{-1}$ and a spectral index of 2.3, at a distance of $100$~Mpc. The contribution of primary protons (green dashed), helium (blue dash dotted) and iron nuclei (red solid) are indicated.}  \label{fig:neut_comp}
\end{flushleft}
\end{figure}

Figure~\ref{fig:neut_comp} shows the contribution of protons, helium and iron to the production of neutrinos. It appears that the contribution of protons dominates by far those of other nuclei, but the different components exhibit the same shape. The gap between the different species is mainly due to the relative abundances we assumed (similar to the Galactic cosmic ray composition) and the different contributions would be closer if all the species had similar abundances. Even in this case however, the contribution of the different species would decrease with mass. This is due to the fact that the energy of the neutrinos depend on the nucleus Lorentz factor $E_\nu\propto\Gamma_A\propto E_A/A$, which means that a neutrino produced at energy $E_\nu$ requires a higher cosmic ray energy $E_A$ if $A$ is large. For the typical spectral index we assume in this study ($\geq 2$) the contribution of light nuclei is larger. The hierarchy between the different primary will of course depend on the spectral index of the source, hard spectra giving a smaller gap between the different species. The lower interaction lengths of heavy nuclei tend to partially counterbalance this effect as heavy nuclei usually suffer more interactions potentially producing neutrinos.

Nuclei can produce neutrinos either by interaction of the parent nucleus (through hadronic interactions or photopion production) or by the interactions of the secondary nucleons (hadronic, photopion or neutron decay but the contribution of the latter is very low in the energy range we consider). The relative contribution of the secondary nucleons ultimately depends on the mass of the primary cosmic ray, as the mean free paths for photo-hadronic and hadronic interactions read $l_{A\gamma}\propto A^{-1}$ and $l_{Ap}\propto A^{-2/3}$. As a result, in the case of heavy nuclei, the secondary nucleons are produced very rapidly, when the nucleus is still in the central regions of the cluster. This is not the case for lighter nuclei such as Helium. The secondary nucleons produced in the central regions of the cluster will encounter denser photon and baryon backgrounds and are therefore more likely to produce neutrinos. The contribution of the secondary nucleons to the neutrino fluxes produced inside the cluster is then higher for the heaviest nuclei. 

The competition between hadronic and photonic interaction is also an interesting aspect of the neutrino production in galaxy clusters. As mentioned before hadronic interactions dominate over photo-interaction for the production of neutrinos (for neutrinos produced below $E_{\nu}=10^{18}$ eV). In the case of protons, the hadronic interaction rate is far higher in the cluster center and photo-interaction partially compensates their deficit due to the flatter evolution of the photon density as a function of the distance to the cluster center. Let us note that below $E=10^{16}$ eV, photo-interaction ceases to contribute to the neutrino production due to the energy threshold of the photopion process with the ambient photon background. In the case of nuclei, the gap between the contribution of hadronic and photo-interaction is larger. This can appear odd at first thought as the rate of photo-interactions $\propto A$ grows faster with the mass than the hadronic interaction rate $\propto A^{2/3}$. Yet a significant fraction of the photo-interactions happens through the giant dipole resonance process which only produces neutrinos via the interactions of secondary nucleons. The photopion production of nuclei that can produce directly neutrinos whenever the produced pion is not absorbed in the nucleus is then screened by this less efficient process. On the other hand, hadronic interactions of the parent nucleus produce neutrinos in all the cases which explains the lower relative contribution of photo-interactions to the neutrino production in the case of nuclei. One may also note, that due to the different evolutions with the distance to the cluster center of the photon and baryon backgrounds the relative contribution of the photo-interactions will be higher when the source is shifted from the cluster center.\\

As an illustration, the sensibility of the next generation detector KM3NeT will be of order $\sim 2\times10^{-9}$~GeV~s$^{-1}$~cm$^{-2}$ for a year for a point source around PeV energies. We may expect from these plots that the flux  from this galaxy cluster is typically below the current observable limits of neutrinos for a spectral index of 2.3. The flux is boosted of one order of magnitude for an index of 2.1 and is closer to experimental sensitivities (see Fig.~\ref{fig:neut}), assuming the same integrated luminosity above 1~GeV. Yet, as pointed out in the previous subsection, such a hard injection spectrum would overproduce ultrahigh energy cosmic rays as compared to the observed data, by a factor of $\sim 50$. This means that this kind of central sources with high luminosities and hard spectral indices, more favorable for the detection of point sources, should have a low density if one does not want to violate the limits imposed by the total cosmic ray flux. One can infer from these results that detections from a single cluster seem difficult for the moment, but there could be some chance of observing diffuse fluxes.

\begin{figure}[tb]
\begin{flushleft}
\includegraphics[width=0.45\textwidth]{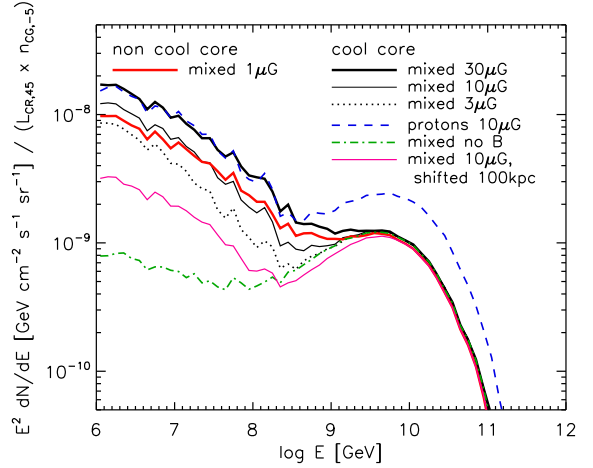} 
\caption{Diffuse neutrino fluxes obtained with galaxy cluster density $n_{\rm s}=10^{-5}$~Mpc$^{-3}$ and AGN cosmic ray luminosity $L_{\rm cr}=10^{45}$~erg~s$^{-1}$. A mixed composition is injected at the center of the non cool core cluster with $B_{\rm c}=1~\mu$G (red thick solid), and in cool core clusters with $B_{\rm c}=30~\mu$G (black thick solid), $B_{\rm c}=10~\mu$G (black thin solid), $B_{\rm c}=3~\mu$G (black dotted) and without magnetic field (green dash dotted). We also present the cases of a pure proton injection at the center (blue long dashed) and a mixed composition injected at 100~kpc from the center of a cool core cluster of  $B_{\rm c}=10~\mu$G (pink solid). }  \label{fig:neutmany}
\end{flushleft}
\end{figure}

In this token, we calculate the expected cumulative neutrino fluxes for various parameter sets, for a galaxy cluster density $n_{\rm s}=10^{-5}$~Mpc$^{-3}$, assuming the same mass for all galaxy clusters, as we did previously for the cosmic ray diffuse flux. Our results are presented in Fig.~\ref{fig:neutmany}. We do not take into account the evolution of the source density and luminosity with redshift which should have a negligible effect for redshifts $z<0.2$ \cite[]{CB98}. For higher redshifts, one should also consider the evolution of the cluster itself (density, infrared background, and most of all magnetic field), which is an impossible task seen our poor knowledge on the origin and evolution of the extragalactic magnetic field. 
Our calculations enables us to capture the essential features due to key parameters, and to notice that all our fluxes lie around the observable threshold of current and upcoming experiments. Indeed, the differential sensibility of IceCube for diffuse fluxes is of order $1.5\times 10^{-8}$~GeV~s$^{-1}$~cm$^{-2}$~sr$^{-1}$ for one year, which leaves room for a positive detection of signals coming from clusters of galaxies around 1~PeV. In this energy range, our fluxes are above the expected cosmogenic neutrino fluxes because of the magnetic confinement and enhanced baryon and photon backgrounds in the cluster environment.

On the contrary for ultrahigh energies, cosmic rays are not significantly confined, hence we find that the neutrinos produced inside the galaxy cluster by interactions with CMB photons (for which we take into account the cosmological evolution) only represent a fraction of the total cosmogenic neutrinos and will thus be dominated by them. 

It is interesting to notice the importance of the magnetic confinement for the production of secondary neutrinos, as well as the differences between cool core and non cool core clusters. As we pointed out in the previous section, one should only compare the non cool core case presented here with the cool core case at $B_{\rm c}=30~\mu$G. It appears that the the presence of the magnetic field enhances the neutrino production of an order of magnitude, but there is only a slight difference between the various magnetic intensity and configurations. We note that a pure proton composition leads to a similar neutrino flux as compared to a mixed Galactic composition. The flux is actually even slightly higher, as protons produce more neutrinos than nuclei (see above) for the choices of astrophysical parameters we have made. In the case of a pure proton composition, steeper source spectral indexes would be required to fit the cosmic ray spectrum, which means that extremely large luminosities for single source could be allowed (even too large to be realistic if one assumes a single power law down to $10^9$ eV). This argument on the luminosities is usually alleviated by invoking a change in the injected spectrum at some energy (see \citealp{BG07} and \citealp{MIN08} in the context of galaxy clusters). As the fluxes again scale with $L_{\rm cr}\times n_{\rm s}$, an increase in one of these parameters could enhance the neutrino rate. One should yet remember that the cosmic ray fluxes would then be overproduced as compared to the observed data, as we calculated in section~\ref{subsection:crspec}. The constraints imposed by the total cosmic ray flux on the diffuse neutrino flux are indeed quite stringent. However the dilution of the flux due to the limited AGN lifetime can be used to justify an increase of the luminosity by a factor $t_{\rm AGN}/t_{\rm cycle}$, where $t_{\rm cycle}$ is the periodic duration of an AGN cycle. 

Our results are  consistent with the analytical treatment of \cite{BBP97} -- and with the study of \cite{MIN08} though they assumed different physical parameters. A rough order of estimate on the neutrino flux $J_\nu$ around PeV energies in the case of a pure proton composition, assuming that the hadronic interactions are the dominant interaction process can be written \cite[]{MIN08} : 
\begin{eqnarray}
E^2J_\nu(E) \sim 0.7 \times {10}^{-11}~{\rm GeV~s^{-1}~{cm}^{-2}}\times \nonumber\\
\left(\frac{f_{\rm pp}}{2.4 \times {10}^{-3}}\right) \, \left(\frac{D}{100~{\rm Mpc}}\right)^{-2} \left(\frac{L_{E,16}^{\rm cr}}{{10}^{43}~\mbox{\ergpers}}\right)\, ,
\end{eqnarray} 
where $D$ is the distance to the source, $L_{E,16}^{\rm cr}=10^{43}$~erg~s$^{-1}$ the cosmic ray luminosity at $E={10}^{16}~{\rm eV}$ (corresponding roughly to a value of $L_{E,16}^{\rm cr}=10^{45}$~erg~s$^{-1}$ for a minimum injection energy of $E_{\rm min}=10^9$~eV, with spectral index 2.3) and $f_{\rm pp}$ the effective optical depth for the proton-proton interactions at energy $E \sim {10}^{16}~{\rm eV}$. This latter quantity can be written: $f_{pp} = 0.8\,\sigma_{pp}n_Nct_{\rm esc}\sim2.4\times 10^{-3}(n_H/10^{-4.5}~{\rm cm^{-3}})(t_{\rm esc}/1~{\rm Gyr})$, assuming a constant baryonic density $n_H$ and escape time $t_{\rm esc}$ throughout the cluster. 

Our fluxes are lower than those calculated by \cite{demarco06} in the energy range range between $10^{16}$ and $10^{18}$ eV, and show an overall difference in the shape of the energy spectrum. This discrepancy stems mainly, as already mentioned in section~\ref{subsection:bg}, from their choice of very bright infrared galaxy SED (instead of elliptical galaxy in the present study) to calculate the cluster photon background. Furthermore, hadronic interactions were not taken into account by \cite{demarco06}. 

The neutrino fluxes presented in Fig.s~\ref{fig:neut}, \ref{fig:neut_comp} and \ref{fig:neutmany} do not take into account the limited AGN lifetime and assume a permanent emission regime. This is justified for the highest energy cosmic rays that produce neutrinos through interactions with the CMB photons and that are not trapped inside the cluster: neutrino production in this case should thus happen quickly after the injection. We checked that it is also the case for the relatively lower energy particles. Indeed, most of the PeV energy neutrino flux is produced in the central region of the cluster shortly after injection.

\subsection{Secondary gamma rays}
Secondary gamma rays can also be a signature of the propagation of ultrahigh energy protons or nuclei in clusters of galaxies. As for the neutrinos, the simultaneous observation of charged particles and of gamma ray photons from a cluster will depend on the duration of the life cycle of the source. 

Very high energy charged and neutral pions are produced via hadronic and photo-hadronic interaction processes. Neutral pions decay into high energy gamma rays, while charged pions lead to high energy electrons and positrons that lose their energy through synchrotron and Inverse Compton emissions. Generated gamma rays are attenuated by the pair-creation process on the intergalactic photonic background during their propagation, while electrons and positrons typically lose most of their energies via synchrotron and Inverse Compton losses locally and sufficiently low energy pairs are trapped by the magnetic fields of clusters.

In our propagation code, the gamma ray, electron and positron production via pion decay is treated in a discrete manner, 
as in \cite{Allard06} for nuclei projectiles with $A> 1$ and with SOPHIA for proton and neutron cosmic rays. Electron and positron pair creation through photo-hadronic processes is implemented as in \cite{ASM06}. For each time step (chosen to be much larger than the typical mean free path of pair photo-production processes), we assume that an ensemble of electron and positron pairs are generated with energies distributed according to a power law, and maximum energy depending on the primary particle. As we are interested in the cascaded photons, the use of an improved pair spectrum as in \cite{KA08} is not expected to change our results.

Electromagnetic cascades are treated as post-analysis, using the radial distribution of electrons, positrons and photons produced in the simulation, inside the cluster of galaxies. Very high energy photons propagate in the direction of ultrahigh energy cosmic rays, while lower energy photons are radiated almost isotropically. Here, for simplicity, we approximate that photons propagate in the radial direction. Although this is a crude approximation, it is justified in so far as the energy attenuation length due to the CMB (at $\sim$~PeV) is much shorter than the cluster size of $\sim$~Mpc, and it is enough to obtain qualitatively appropriate results (Asano 2009, private communication).The resulting high energy gamma ray fluxes are calculated by solving kinetic equations (see for example \citealp{Aharonian02,Murase09} and references there in). 
 
\begin{figure}[tb]
\begin{center}
\includegraphics[width=0.45\textwidth]{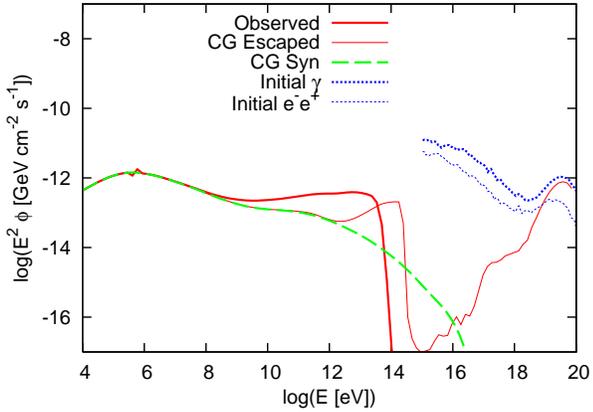} 
\caption{\footnotesize Gamma ray emission from a source placed in a cool core galaxy cluster with  $B_{\rm c} = 3~\mu$G, a luminosity of $L_{\rm cr} = 10^{45}~$\ergpers, with spectral index 2.3, and located at a distance of 100~Mpc. Blue dotted lines represent electron and positron (thin) and photon (thick) fluxes produced in the cluster by cosmic rays of energy $E\ge E_{\rm min, simu}=10^{16}$~eV. Green dashed lines indicate the contribution of synchrotron emission that occurred in the cluster. The red thin line is the gamma ray flux obtained after propagation in the cluster of galaxies, that one would observe at a distance of 100~Mpc in absence of electromagnetic cascades in the extra-cluster medium. The thick red line indicates the photon flux observed at 100~Mpc after electromagnetic cascades, assuming an extra-cluster average magnetic field intensity of $B_{\rm IG}=10^{-13}$~G.}  \label{fig:gamma1}
\end{center}
\end{figure}

\begin{figure}[tb]
\begin{center}
\includegraphics[width=0.45\textwidth]{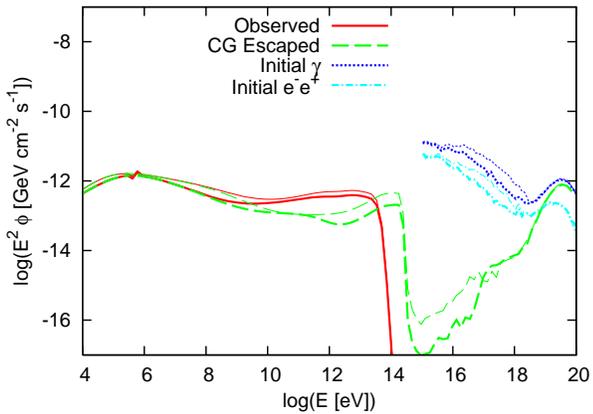} 
\caption{\footnotesize Comparison between the fluxes obtained for a cool core cluster with $B_{\rm c}=3~\mu$G  (thick lines) and a non cool core cluster with $B_{\rm c}=1~\mu$G (thin lines) -- the other parameters are identical to those of Fig.~\ref{fig:gamma1}. Green lines represent the flux obtained after propagation in the cluster, as one would observe it at 100~Mpc in absence of electromagnetic cascades in the extra-cluster medium. Red lines give the gamma ray flux observed at 100~Mpc after electromagnetic cascades, assuming an extra-cluster average magnetic field intensity of $B_{\rm IG}=10^{-13}$~G.}  \label{fig:gamma2}
\end{center}
\end{figure}

In Figure~\ref{fig:gamma1}, we show the spectra of produced gamma rays from a cool core cluster with $B_{\rm c}=3~\mu$G, which is located at a distance of 100~Mpc, with a source of luminosity $L_{\rm cr}=10^{45}$~erg~s$^{-1}$ and of spectral index of 2.3 (the parameters here are identical to those used for neutrino fluxes in Fig.~\ref{fig:neut}). The case of a non cool core cluster with $B_{\rm c}=1~\mu$G is also shown in Fig.~\ref{fig:gamma2}. Only the contribution of cosmic rays with $E \gtrsim {10}^{16}~{\rm eV}$ is shown in these plots. Very high energy electrons mainly radiate their energy through synchrotron emission and lead to the production of gamma rays around keV-GeV energies. On the other hand, gamma rays produced via Inverse Compton emission are expected above the TeV range, which is indeed observed in the figures (thin red line in Fig.~\ref{fig:gamma1} and green dashed lines in Fig.~\ref{fig:gamma2}). 

Thick red line in Figure~\ref{fig:gamma1} and red lines in Figure~\ref{fig:gamma2} indicate the flux obtained after propagation of photons, electrons and positrons in the extra-cluster medium. Gamma rays with $\gtrsim$~PeV gamma rays are significantly attenuated by the cluster photon field and the CMB (where the cluster photon field itself causes only a minor effect). Above $\sim 10$~TeV, photons also suffer from interactions with the diffuse infrared, microwave and radio backgrounds. Thus, they are largely absorbed and cascaded into lower energy gamma rays especially for high redshift clusters of galaxies. 
In our calculations, we set the intergalactic magnetic field strength to $B_{\rm IG}={10}^{-13}$~G, which allows us to consider the cascades as unidimensional as long as the coherent length is short enough so that the deflection angle is small, $\theta_B \ll 1$ \citep{Murase08}. For stronger magnetic fields or longer coherent lengths, one must consider the deflections experienced by electrons and positrons, which will lead to a dilution of the flux. The synchrotron emission from electrons and positrons near the source (green dashed line in Fig.~\ref{fig:gamma1}) as well as photons that are directly produced by cosmic rays below some TeV will not be affected, as the Universe appears transparent to those particles. It should be noted however that the contribution of photons produced by cascades in the intergalactic medium is relatively limited below $\sim 1$~TeV as compared to the flux produced in the cluster itself (see Fig.~\ref{fig:gamma1}).

We also assume that our clusters of galaxies harbour a steady AGN as a central source. Note that considering a finite AGN activity time could lead to a diminished gamma ray flux around TeV energies, if the intergalactic magnetic field is strong enough to spread the cascaded gamma ray emission over a much longer time than the AGN lifetime. These issues will eventually depend on the rate of magnetic enrichment of the intergalactic medium. As for the synchrotron emission directly produced in the cluster, and that constitutes the major gamma ray contribution at lower energies, a finite AGN lifetime could lead to a complete temporal decorrelation of the gamma ray flux from the cosmic ray flux, according to the ratio between the source activity and inactivity durations.

We observe in Fig.~\ref{fig:gamma1} and \ref{fig:gamma2} the same effects as for neutrinos, depending on the configuration and the intensity of the magnetic fields. For gamma rays, one may notice an additional contribution of the intensity of the field on the synchrotron emission: for stronger fields, electrons and positrons produce a larger amount of photons of higher energy (the energy loss length of an electron of energy $E_e$ is indeed $x_{\rm syn}\propto B^{-2}E_e^{-1}$ and the synchrotron emission peaks at energy $E_{\gamma,\rm syn}\propto BE_e^2$). These features are not striking in figure~\ref{fig:gamma2}, where the magnetic field is always strong enough to have an important synchrotron emission. The scenario in which the cluster magnetic field is set to zero suppresses completely the flux below $\sim 10$~GeV. 

\begin{figure}[tb]
\begin{center}
\includegraphics[width=0.45\textwidth]{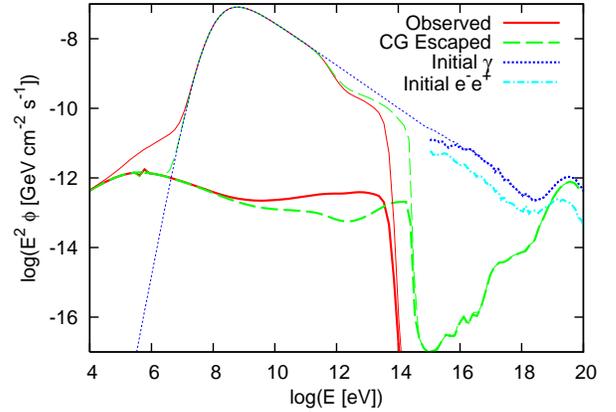} 
\caption{\footnotesize Contribution to the gamma ray flux of low energy cosmic rays with $E<E_{\rm min,simu}=10^{16}~$eV, for a cool core cluster with $B_{\rm c}=3~\mu$G  -- the other parameters are identical to those of Fig.~\ref{fig:gamma1}. Thick lines represent the calculated emission for cosmic rays of energy $E>E_{\rm min,simu}$, and the thin lines the emission including low energy cosmic rays modeled semi-analytically in the diffusive approximation. The cosmic ray injection time is set to $3$~Gyr, assuming that the sources emit particles almost continuously.}  \label{fig:gamma_low}
\end{center}
\end{figure}

\begin{figure}[tb]
\begin{center}
\includegraphics[width=0.31\textwidth,angle=-90]{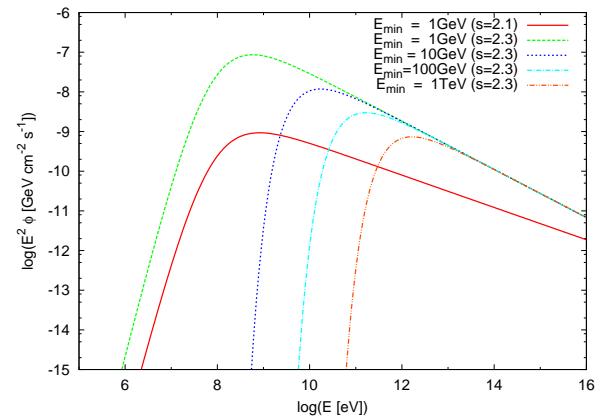} 
\caption{\footnotesize For illustrative purposes: gamma ray fluxes produced by low energy cosmic rays ($E<E_{\rm min,simu}$) assuming spectral indices of 2.1 (red solid) and 2.3, with different minimum injection energy $E_{\rm min}=1,10,100,10^3$~GeV. The source luminosity is assumed to be $L_{\rm cr}=10^{45}$~\ergpers{} for spectral index 2.3, and $L_{\rm cr}=2\times 10^{43}$~\ergpers{} for spectral index 2.1 between 1~GeV and $10^{20.5}$~eV.}  \label{fig:gamma_lowcomp}
\end{center}
\end{figure}

We examine the contribution of cosmic rays of energy $E < E_{\rm{min,simu}}=10^{16}$~eV to the flux of gamma rays in the GeV-TeV energy range. The propagation of these low energy particles has not been calculated numerically because of the excessive calculation time that it implies. It is however possible to calculate the spatial distribution and the energy spectrum of these cosmic rays, assuming that they are in a diffusive regime, which is appropriate in this energy range. As gamma rays are mainly produced by protons through hadronic interactions in this energy range (the discussion is similar to the case of neutrinos, see previous section), we consider only proton-proton interactions in our calculations. We assume an initial Gaussian distribution that extends over $\sim 1$~kpc in the cluster, and we further suppose that the central source injects particles at a rate that maintains this gaussian distribution at all times. The diffusion equation is solved by using a Crank-Nicholson scheme, adopting the following diffusion coefficient computed according to the results of \cite{CLP02} : $D=1.2\, r_{\rm L} c \left[ r_{\rm L}/l_{\rm c} + 0.1 (r_{\rm L}/l_{\rm c})^{-2/3} \right]$, where $r_{\rm L}$ is a Larmor radius and $l_{\rm c}$ is a coherence length of the magnetic field \cite[][Appendix A]{KL08b}. 

In the inner region of the cluster ($\sim 10$~kpc), we obtain an energy spectrum steeper than the injected one, as low energy particles are accumulated while higher energy particles escape from the inner region earlier. The opposite effect is observed in the outer regions ($\sim 1$~Mpc), where low energy particles are less numerous. One can note that the calculated initial photon spectrum slope is of order $-2.6$, which corresponds as expected to the combination of the injection slope ($-2.3$) and to the evolution of the confinement time with energy ($-1/3$). We evaluate the gamma ray flux produced by proton-proton interactions through the formulations of \cite{PE04} and \cite{AN08}.  Finally, we match the resulting power-law spectrum with the numerically calculated one by interpolating the value between $E=8\times10^{15}$~eV (corresponding to the maximum value of our calculation under the diffusion approximation) and $E=10^{16}$~eV (corresponding to the minimum value of the higher energy spectrum). Such a rough matching is sufficient for demonstrative purpose. 

Our results are shown in Fig.~\ref{fig:gamma_low} and \ref{fig:gamma_lowcomp}, which show that gamma ray signatures of ultrahigh energy cosmic rays are overwhelmed by gamma rays produced by lower energy cosmic rays. The situation looks less critical for harder cosmic ray spectra, but the photons emitted by ultrahigh energy cosmic rays will still be one or two orders of magnitudes below the low energy particle contribution for spectral indices of $2.0-2.1$. \\

We discuss in what follows the detectability of photons from clusters of galaxies. The future Cherenkov Telescope Array (CTA) is expected to have a sensitivity of order $\sim 10^{-11}$~GeV~$\rm s^{-1} cm^{-2}$ for 100 hours for a point source around TeV energies. For an extended source, the sensitivity becomes weaker by a factor of $\theta/\theta_{\rm{PSF}}$, where $\theta$ and $\theta_{\rm{PSF}}$ are the angular extensions of the source and of the point spread function respectively. The latter being of order of a fraction of arcminute around 1~TeV, we might assume that a typical cluster of galaxy observed at a distance of 100~Mpc leads to an angular size of $\theta \sim 10  \,\theta_{\rm{PSF}}-100 \,\theta_{\rm{PSF}}$, which corresponds roughly to a sensitivity of $\sim 10^{-10}-10^{-9}$~GeV~$\rm s^{-1} cm^{-2}$. The Fermi satellite has a lower angular resolution (of a fraction of degree) around GeV energies, leading to a sensitivity reaching $\sim 2 \times 10^{-10}$~GeV~$\rm s^{-1} cm^{-2}$ for a year for a typical cluster of galaxies at 100~Mpc in this energy range.

Figure~\ref{fig:gamma1} shows that the gamma ray signal coming from ultrahigh energy cosmic rays would be hardly detectable by any of these instruments. This conclusion also applies to the future detectors HESS-2 and MAGIC-2. Those signals might moreover be overwhelmed by a stronger contribution of low energy cosmic rays, as pointed out in Fig.~\ref{fig:gamma_low}. 

The gamma ray flux from cosmic rays of low energy might be detected by the experiments cited above, but they should first be compared to existing observational limits. The EGRET point source limit is of order $3-4\times 10^{-9}$~\gevperspercm{} at 100~MeV, for a cluster like Coma located at $\sim100$~Mpc \citep{PE04}. We thus note that the values presented in Fig.~\ref{fig:gamma_lowcomp} for spectral index 2.3 violate this limit in the energy range $\sim10^{8-10}$~eV. Besides, an estimate of the diffuse gamma ray flux using the same method as for neutrinos leads to a value of $\sim 5\times10^{-5}$~\gevperspercm{}~sr$^{-1}$ at 1~GeV for a spectral index of 2.3, which is also above the diffuse EGRET limit \citep{Osborne94}. 

It might be reminded however, that these plots were made under the assumption of a stationary cosmic injection regime, and that considering a limited AGN lifetime and particular configurations of sources may lower enough the flux to avoid the violation of the EGRET limit. These problems can also be overcome in the case of a harder cosmic ray injection spectrum ($\le 2.1$), that can be either a continuous power law extending to ultrahigh energies (in which case one should remember that the luminosity or the density of the source should be lowered by a factor $\sim 50$ in order not to overshoot the observed cosmic ray data, implying a luminosity of $L_{\rm cr}=2\times10^{43}$~\ergpers{} between $E=10^9-10^{20.5}$~eV), or a broken power law. In the case of a 2.1 low energy injection spectrum, our calculated diffuse flux from clusters of galaxies would contribute to $\sim 20\%$ of the EGRET diffuse flux. Such a flux should be also detectable by Fermi around GeV energies. In any case, future observations will be able to put constraints on our predictions.

We also made for these calculations the assumption that AGN in clusters of galaxies were the major contributors to the total ultrahigh energy cosmic ray flux in order to set the source luminosity and density. One may however alleviate this hypothesis and invoke a lower source density or luminosity leading to a better consistency with the observed limits. Moreover, these results raise the question of the minimum energy of cosmic rays injected in the cluster of galaxies. It is indeed not obvious that cosmic rays can escape the acceleration site of AGN at an energy as low as 1~GeV, due for example to high magnetic confinement times and shorter adiabatic energy loss times \cite[]{BGG06}. If the cosmic ray injection starts at an energy greater than 100~GeV (as presented in Fig.~\ref{fig:gamma_lowcomp}), the fluxes could be compatible with observed data, even with spectral indices of 2.3, but their detectability by current and upcoming instruments might be compromised.  

Ultrahigh energy photons could also be a potential probe of ultrahigh energy cosmic rays \cite[]{Murase09} and they could even produce a pair halo after they leave a cluster. However, our results imply that their detection by current and future detectors such as CTA does not seem easy as well, unless the source is rather close or luminous.

In this paper, we did not evaluate the energy spectra of very low energy photons, especially in the radio band. Values for the diffuse radio emission flux from some clusters of galaxies were reported by several authors \citep{Giovannini93,GF00,Feretti04, Bagchi06,FG08,Brunetti08}. In a few years, the Low Frequency Array (LOFAR) and the Long Wave length Array (LWA) will even observe clusters at low radio frequencies, and possibly discover the bulk of the cluster scale synchrotron emission in the Universe \citep{ER02,Brunetti08}. 

Finally, it should be noted that these low energy photons as well as $\lesssim$~TeV gamma rays may come from leptonic components \citep[e.g.,][and references therein]{EB98, AV00, TK00, KW09} rather than hadronic ones. One promising way of distinguishing them would be to observe the spatial distribution of gamma rays, since hadronic gamma rays trace the gas while leptonic ones trace the shocks \cite[]{Miniati03}.

\section{Conclusion}
We studied the propagation of nuclei in a magnetised cluster of galaxies using a complete propagation code and based on a detailed study of the physical properties of clusters of galaxies. Due to their strong magnetic fields and dense matter and radiation backgrounds in the central regions, galaxy clusters are in principle hostile environments for cosmic ray nuclei.
It is found that the nuclei survival depends mostly on the strength and the profile of the magnetic field, as well as on the position of the source. Indeed, heavy nuclei can be efficiently suppressed when the source is found at the very center of a cool core cluster for the strongest magnetic field normalizations. Light and intermediate nuclei are found to be much less affected by the confinement and the dense central photon and baryon backgrounds even when strong magnetic fields are at play. In the case of lower central magnetic fields or if the source is shifted from the central regions, the survival of the heaviest elements becomes easier.

We also examined the effect of the limited lifetime of the central source. The resulting spectrum from a single source is found to be time dependent due to the rigidity dependence of the confinement time in the magnetized regions. As a consequence, at a given energy, the composition becomes heavier, the flux fainter with time and the overall spectrum gets softer. These effects, though expected for any source surrounded by a magnetized region, should not affect the overall cosmic ray spectrum and composition providing that a sufficiently large number of source is present within the cosmic ray horizon at a given energy. Let us note that depending on the duration of the life cycle, one can observe preferentially the contribution of both cosmic rays and secondary neutrinos and gamma rays, or the former without the latter, or even the latter without the former. 

Concerning secondary neutrinos, we have shown that their production can be significantly enhanced by the confinement of the cosmic rays (below $\sim Z \times 10^{18}$~eV) in the central regions of the cluster.  As a consequence,  neutrinos below $10^{17}$~eV, produced mainly via hadronic interactions or interactions with ambient infrared and optical photons, could form a detectable diffuse flux for IceCube or KM3Net if a significant fraction of all the UHE cosmic rays originates from the central regions of galaxy clusters. At the highest energies however, cosmic rays are barely confined by the cluster magnetic fields and as a result, the diffuse neutrino flux from galaxy cluster above $10^{18}$ eV should only represent a small fraction of the whole diffuse cosmogenic neutrino flux, preventing the signature of their origin. The flux from single sources (except if they are exceptionally bright) should be well below the current observation capabilities. 

Gamma rays are another potential signature of the presence of cosmic ray accelerators in Galaxy clusters. We find that the fluxes produced by very and ultrahigh energy cosmic rays should remain typically below the reach of current and planed experiments for reasonable assumptions on the sources luminosity, and corresponding source densities of $n_{\rm s} \gtrsim {10}^{-6}~{\rm Mpc}^{-3}$. Moreover, due to the combined effects of the cascading of the produced photons and the evolution of the confinement time with the rigidity (coupled to the flatness of the hadronic interaction cross sections), this contribution is likely to be overwhelmed by the one of lower energy cosmic rays. If the latter component is released in the intra-cluster medium, ample gamma ray fluxes, testable by satellites or ground based experiments could be produced.

As a conclusion, multi-messenger observations could be used to test the presence of cosmic ray accelerators in galaxy clusters. Though ultrahigh energy cosmic rays are more likely to be probed by their arrival directions, very high and low energy cosmic rays could be revealed using neutrino and gamma ray observations.

\acknowledgments
We thank Katsuaki Asano, Klaus Dolag, Herv\'e Dole, Torsten En{\ss}lin, Martin Lemoine, Christophe Pichon, Thierry Sousbie, and Hajime Takami for interesting discussions. K.K. was partly supported by the FY2008 Japan Society for the Promotion of Science (JSPS) Postdoctoral Fellowship for this work.  S.N. was supported by Grant-in-Aid for Scientific Research on Priority Areas No.~19047004 by Ministry of Education, Culture, Sports, Science and Technology (MEXT), Grant-in-Aid for Scientific Research (S) No.~19104006 by JSPS, Grant-in-Aid for young Scientists (B) No.~19740139 by JSPS. J.A., K.M. and S.N. are supported by Grant-in-Aid for the Global COE Program "The Next Generation of Physics, Spun from Universality and Emergence" from MEXT of Japan.

\bibliography{cluster}

\clearpage

\end{document}